\documentclass{aa} 

\usepackage{graphicx}
\usepackage{subcaption}
\usepackage{mwe}
\usepackage{natbib}
\usepackage{amsmath}	
\usepackage{amssymb}	
\usepackage{amssymb}
\usepackage{txfonts}
\usepackage{slashed}
\usepackage{gensymb}
\usepackage{empheq}
\usepackage{tabularx}
\usepackage{longtable}
\usepackage{graphicx}
\usepackage[labelfont=bf]{caption}
\usepackage{lmodern}
\usepackage{microtype}
\usepackage{fancyhdr}
\usepackage{rotating}
\usepackage{lscape}
\usepackage{hyperref}
\hypersetup{colorlinks=true,citecolor=blue,linkcolor=blue,urlcolor=blue}
\newcommand*\disperse{D{\scriptsize IS}P{\scriptsize ER}SE}

\begin{document}

   \title{Pre-processing of galaxies in cosmic filaments around AMASCFI clusters in the CFHTLS}
   
\titlerunning{Cosmic filaments in the CFHTLS}
\authorrunning{Sarron et al.}

\author{F. Sarron\inst{1},
        C. Adami\inst{2},
        F. Durret\inst{1},  
           and C. Laigle \inst{3}
 \offprints{Florian Sarron,  \email{sarron@iap.fr}}
}
\institute{Sorbonne Universit\'e, CNRS, UMR 7095, Institut d'Astrophysique de Paris, 
                98bis Bd Arago, 75014 Paris, France\\
              \email{florian.sarron@iap.fr}
         \and
          Aix Marseille Universit\'e, CNRS, Laboratoire d'Astrophysique de
          Marseille, Marseille, France
          \and
          Sub-department of Astrophysics, University of Oxford, Keble
          Road, Oxford OX1 3RH
             }

   \date{Received March 4th, 2019, ; accepted Month Day, Year}

 
  \abstract
   {Galaxy clusters and groups are thought to accrete material along the preferred direction of cosmic filaments. Yet these structures have proven difficult to detect due to their low contrast with few studies focusing on cluster infall regions.}
   {In this work, we detected cosmic filaments around galaxy clusters using
     photometric redshifts in the range $0.15 < z < 0.7$. We characterised
     galaxy populations in these structures to study the influence of ``pre-processing'' by cosmic filaments and galaxy groups on star-formation quenching.}
   {The cosmic filament detection was performed using the {\small AMASCFI} Canada-France-Hawaii Telescope Legacy Survey
     (CFHTLS) T0007 cluster sample (Sarron et al. 2018). The filament reconstruction was done with
     the \disperse~algorithm in photometric redshift slices. We showed that this reconstruction is reliable for a CFHTLS-like survey at $0.15 < z < 0.7$ using a mock galaxy catalogue.
     We split our galaxy catalogue in two populations (passive and star-forming) using the LePhare SED fitting algorithm and worked with two redshift bins ($0.15 < z \le 0.4$ and $0.4 < z < 0.7$).}
   {We showed that the {\small
       AMASCFI} cluster connectivity (i.e. the number of filaments connecting to a cluster) increases with cluster mass $M_{200}$. 
     Filament galaxies outside $R_{200}$ are found to be closer to clusters at low redshift, whatever the galaxy type. Passive galaxies in filaments are closer to clusters than star-forming galaxies in the low redshift bin only.
     The passive fraction of galaxies decreases with increasing clustercentric distance up to $d \sim 5$ cMpc. Galaxy groups/clusters that are not located at nodes of our reconstruction
   are mainly found inside cosmic filaments.}
   {These results give clues for ``pre-processing'' in
     cosmic filaments, that could be due to smaller galaxy groups. This trend could be further explored by applying this method to larger photometric surveys such as {\small HSC-SPP} or {\it Euclid}.}


   \maketitle
%
\section{Introduction}\label{sec:web-context}
Matter in the universe is not distributed uniformly but rather tends
to aggregate into a complex structure with rich and poor galaxy
clusters connected by filaments and sheets
surrounding regions almost devoid of galaxies - cosmic voids. This
network of structures forms the so-called cosmic web.\\
\indent This structure was first observed using spectroscopic redshifts with
the pioneering work of \citet{deLapparent1986}. Since then, this
galaxy distribution has been observed in great detail by many surveys, either shallow (so limited to low redshifts) but on large
portions of the sky (\citealp[e.g. 2 degree Field Galaxy Redshift Survey (2dRGRS)][]{Colless2001}; \citealp[and
Sloan Digital Sky Survey (SDSS)][]{York2000}) or deeper (so probing higher redshifts) but limited
to smaller regions \citep[e.g. VIMOS VLT Deep Survey (VVDS)][]{VVDS05}. Recent efforts to probe higher redshifts on significant areas of the sky have been made
\citep[e.g. VIMOS Public Extragalactic Redshift Survey (VIPERS)][]{Guzzo2014}.\\
\indent Numerical simulations of dark matter particles
\citep[e.g][]{Springel05} have led to similar results. These
simulations allow to grasp the dynamical aspect of the formation and
evolution of these structures. Dark matter is shown to aggregate in a
bottom-up fashion forming bigger and bigger structures through cosmic
time (starting with galaxies, up to rich clusters). In this process,
matter is expelled from the voids and aligns into the sheets/walls, where it
gets accreted in filaments. If clusters at the nodes of the cosmic web
are believed to form mostly at $z > 1$, they keep accreting galaxies
along the preferential direction of the filaments they are connected
to at lower redshift \citep{Bond1996}.

These galaxy clusters host a population
of quiescent galaxies (the so-called red sequence) that formed at $z >
1$
\citep[e.g.][]{Mullis2005,Mei2006,Eisenhardt+08,Kurk2009,Hilton2009,Papovich2010}
and keep being enriched at lower redshift
\citep[e.g.][]{Rudnick+09,Zhang+17,Martinet+17,Sarron2018}. The
formation and evolution of such a ``red and dead'' galaxy population
implies there are some physical processes at play in quenching
star-formation in galaxies.

Yet, the mechanisms responsible for this quenching  are still poorly
constrained. Indeed, if it is now well established that 
the environment density plays a role in star-formation quenching
\citep[the fraction of quiescent galaxies steadily increases with
environment density,
e.g.][]{Baldry2004,Bamford2009,Peng+10,Moutard2018}, the efficiency of quenching in mildly dense
environments (groups or filaments) is still
unclear, as well as whether or not the high quiescent
fraction observed in clusters is due to specific physical processes in
these environments.
Indeed in the hierarchical structure formation paradigm described above, galaxies
are found to first cluster in small groups inside the filaments that
later collapse into massive clusters \citep[e.g.][]{Contini2016}.

In this context, groups and filaments could be favourable environments
for galaxies to be quenched before entering clusters, a phenomenon
referred as ``pre-processing'' \citep{Fujita2004}. This is a vibrant
topic in the field of galaxy evolution, as understanding where
environmental quenching occurs is crucial to pin-point the physical
mechanisms responsible for it.

``Pre-processing'' by galaxy groups has been extensively studied in
recent years, both based on numerical simulations
\citep[e.g.][]{DeLucia2012,Taranu2014} and
observations \citep[e.g.][]{Smith2012,Roberts2017,Bianconi2018,Olave-Rojas2018}.
The specific role of cosmic filaments in pre-processing started being explored more
recently using spectroscopic surveys, with various teams reporting a
colour/type gradient of galaxies towards filaments
\citep[e.g.][]{Martinez2016,Malavasi2017,Chen2017,Kuutma2017,Kraljic2018}.

These reconstructions of the cosmic web are all based on
  spectroscopic redshifts, which allow to trace filaments in three
  dimensions. Whether one can trace the effect of filaments from
  photometric redshifts is less clear. 
\citet{Malavasi2016} explored how the
  photo-z error impacts the ability to assign the correct environment
  densities to galaxies. They concluded that an uncertainty $\sigma_z \lesssim 0.01 \times
  (1+z)$ provides good environment reconstruction.
  This was confirmed by
  \citet{Laigle2018}, who showed that such a reconstruction is possible
  with very good photometric redshifts ($\sigma_z \sim 0.008 \times
  (1+z)$) by recovering similar stellar mass and colour-type gradients as
  previously mentioned studies \cite[e.g.][]{Malavasi2017} in the redshift range $0.5 < z < 0.9$.

In this work, we first explore the ability of such a method to recover filaments in
the infall regions of clusters based on the less accurate ($\sigma_z \sim 0.03 \times
  (1+z)$) photometric redshifts of the Canada France Hawaii Telescope Legacy Survey (CFHTLS). We aim at understanding whether the environmental quenching of faint
galaxies observed in \citet{Sarron2018} happened inside the cluster region, or if these galaxies have been
pre-processed before entering clusters. To do this, we use our cosmic filament reconstruction to study galaxies located in the cosmic filaments that are feeding galaxy clusters, a question relatively unexplored so far. \citet{Martinez2016} studied filaments between galaxy groups in the SDSS up to $z = 0.15$. They found in particular that filaments play a specific role on quenching galaxies when compared to isotropic infall onto the groups. The method was very recently applied to VIPERS data in the redshift range $0.43 \le z \le 0.89$ \citep{Salerno2019}, where similar results were found, confirming the important role played by filaments in quenching up to $z \sim 0.9$. \citet{Fordinprep} also studied the role of
filaments feeding galaxy groups in the Cosmic Evolution Survey (COSMOS) \citep{Scoville2007} up to $z = 1$ by studying how group properties depend on cluster connectivity (i.e. the number of connected cosmic filaments).\\

In this paper we focus on the study of filaments in the infall regions of clusters in the CFHTLS survey up to $z = 0.7$ based on the AMASCFI cluster catalogue from
\citet{Sarron2018} and on the detection of filaments based on photometric redshifts
with a method adapted from \citet{Laigle2018}. In
Sect.~\ref{sec:data_method} we present our data and our method
to detect cosmic filaments. In Sect.~\ref{sec:validation} we quantify the ability of
our method to recover 3D cosmic filaments, particularly in the infall
region of clusters, using mock data. The method is then applied to the
CFHTLS T0007 data to study quenching in the filaments feeding {\small
  AMASCFI} clusters in Sect.~\ref{sec:CFHTLS_results}. The results are
discussed in Sect.~\ref{sec:discussion}.
We use AB magnitudes throughout the paper, and assume a flat
$\Lambda$CDM cosmology with $\Omega_{\rm M}=0.3$ and $h=0.7$.

\section{Data sets and method}\label{sec:data_method}

Before focusing on the distribution of galaxies in and around the
projected 2D cosmic web, let us first describe the data in hand: the
CFHTLS T0007 and mock data taken from the lightcones of \citet{Merson13}, as well as our method
to reconstruct the cosmic web in these data sets.

\subsection{CFHTLS and mock data}

\subsubsection{CFHTLS T0007}

The photo-$z$ catalogue is obtained from the CFHTLS data release T0007\footnote{available at \url{http://cesam.lam.fr/cfhtls-zphots/files/cfhtls_wide_T007_v1.2_Oct2012.pdf}}. CFHTLS T0007 photo-$z$s were computed in the 154 deg$^{2}$ sky coverage of the CFHTLS from multicolour images in the $u^{*}g'r'i'z'$ filters of MegaCam at CFHT. \\ 
\indent The photo-$z$s were obtained with the LePhare software \citep[]{LePhare99, Ilbert06}. Details on the method are given in \citet{Coupon09}. Briefly, the photo-$z$s were computed using 62 templates obtained after having optimized four templates from \citet{CWW} and two starburst templates from \citet{KinneySB}, and linearly interpolated between them to better sample the colour-redshift space using the VVDS spectroscopic sample \citep[e.g.][]{VVDS05}. A particularly crucial step of the process is the calibration of the zero-points using spectroscopic samples which help in removing biases. The resulting statistical errors on photo-$z$s depend on the redshifts and magnitudes of the galaxies.\\
\indent Following the photo-$z$ catalogue based on the CFHTLS T0007 data release, we define the dispersion as 
\begin{equation}
\sigma _{\Delta z_{\rm phot} / (1 + z_{\rm s})} = 1.48 \times \mathrm{median} \, \left ( \frac{\left | \Delta z  \right |}{(1+z_{\rm s})} \right), 
\end{equation}
\noindent which is the NMAD (Normalized Median Absolute Deviation) estimator defined in \citet{Ilbert06}, with $\Delta z_{\rm phot} = z_{\rm phot} - z_{\rm s}$, where $z_{\rm phot}$ and $z_{\rm s}$  are the photometric and spectroscopic redshifts respectively. The catastrophic failure rate $\eta$ is set as the proportion of objects with $\left | \Delta z  \right | \geq 0.15 \times (1+z_{\rm s})$. \\

\indent In our analysis, we only consider galaxies that are outside the masks provided with the CFHTLS T0007 release. These masks are located around bright stars or artefacts, and mark regions of lower photometric quality. Thus photo-$z$s in these regions would be of poorer quality than those outside the masked regions.

To estimate the photo-$z$ uncertainty at a given magnitude and redshift, we used the median of the errors obtained using SED-fitting binned in magnitude and redshift\footnote{These errors were shown to reliable in the T0007 photo-$z$ release document}.
This is useful to compare the photo-$z$ uncertainties of galaxies of same
absolute magnitude at different redshifts, and thus to have slices
encompassing the same galaxy population at every redshift.

We considered a \citet{BC03} single stellar population calibrated
with the field galaxy luminosity function (GLF) of \citet{Ramos2011} to compute the
apparent GLF knee magnitude at redshift $z$: $m^*_i (z)$. This allowed
us to obtain the redshift evolution of the photo-$z$ uncertainty at
{\it fixed absolute magnitude}. We used this
information to choose the photo-$z$ slice thickness in the CFHTLS (see
Sect.~\ref{sec:method_2D}).\\

\subsubsection{Mock data}

To quantify the quality of the cosmic-web reconstruction from
photometric redshifts, we used a modified lightcone based on a 100 deg$^2$ Deep EUCLID lightcone\footnote{\url{http://astro.dur.ac.uk/~d40qra/lightcones/EUCLID/EUCLID_100_Hband_DEEP.lightcone.tar.gz}} produced following
\citet{Merson13}. The number counts in this lightcone are $\sim 25\%$ lower compared to the CFHTLS T0007 data at all apparent magnitudes ($m_i < m^* + 1.5$). This might impact the contrast of cosmic web structures, including cosmic filaments, in the mock compared to the data. However, this should not be an issue as the AMASCFI cluster finder has been shown to behave similarly in this mock and in the CFHTLS T0007 data \citep{SarronPhD2018}.


We converted the SDSS $i$ band magnitudes of the mock to obtain the
CFHTLS Megcam $i$ band magnitudes as in \citet{Sarron2018} following :
\begin{equation}
i_{\mathrm{Megacam}} = i_{\mathrm{SDSS}} - 0.085 \times (r_{\mathrm{SDSS}}- i_{\mathrm{SDSS}}).
\end{equation}

We added realistic noise to the redshift of each galaxy in the mock
using the median CFHTLS T0007 uncertainty computed in bins of
magnitude and redshift. We refer to \citet{Sarron2018} for a detailed
description of the procedure.

From this information, we computed a redshift probability
distribution function (PDZ) $P_{\rm g}(z)$ for each galaxy in the
mock. This is done following the formalism presented in \citet{CBprobamem}:
\begin{equation}
P_{\rm g}(z) = \frac{1}{\sigma(z,m_{\rm g})} \ \mathrm{exp} \left [ - \frac{(z -
    z_{\rm p,g})^2}{2
  \sigma(z,m_{\rm g})^2} \right ],
\end{equation}
\noindent where $m_{\rm g}$, $z_{\rm p,g}$ and $\sigma(z,m_{\rm g})$ are respectively the magnitude,
photo-$z$ and photo-$z$ uncertainty of the galaxy. While
\citet{CBprobamem} took the simplified prescription
\begin{equation}
\sigma(z,m_{\rm g}) \sim \sigma(z) = \sigma_0 (1+z),
\end{equation}
\noindent here we use our discrete sampling of $\sigma(z,m_{\rm g})$. So, for a
given galaxy, $m_{\rm g}$ and $z_{p,g}$ are fixed, but $\sigma(z,m_{\rm g})$ is a
function of $z$, so that the PDZ will not simply be a Gaussian
centred at $z_{\rm p,g}$. We refer to \citet{CBprobamem} for a more
detailed discussion on this topic.

Finally, as masks due to bright stars in the CFHTLS T0007 catalogue might impact the quality of our reconstruction of cosmic filaments, we added masks representative of the CFHTLS T0007 masks on the mock to make it more realistic. Thus, the effect of the masks are directly included in our validation analysis.

\subsection{Cluster catalogue}

To study cosmic filaments feeding galaxy clusters in the CFHTLS, we
considered the cluster catalogue from \citet{Sarron2018}. This
catalogue was obtained by running the Adami, MAzure and Sarron Cluster
FInder ({\small AMASCFI}) algorithm on the CFHTLS T0007 data. We refer to \citet{Sarron2018} for a full description of the algorithm as well as for detailed properties of the cluster catalogue.

Briefly, candidate clusters are detected based on photometric redshifts
by cutting the galaxy catalogue in redshift slices and detecting peaks
in two-dimensional density maps in each slice. Individual detections
closer than 1 Mpc on the sky and $\Delta z = 0.06$ are then merged
through a Minimal Spanning Tree \citep[MST, see e.g.][]{A10}.

\citet{Sarron2018} computed the selection function of the algorithm in the
CFHTLS. The mean purity and completeness at $z < 0.7$ and cluster
mass $M_{200} > 10^{14}\ {\rm M_\odot}$ are $\sim 90\%$ and $\sim 70\%$ respectively.

Each candidate cluster in the final catalogue has a position, a
photometric redshift (with uncertainty $\sigma_z = 0.018 \times
(1+z)$), a richness estimate and a mass estimate ($M_{200}$). Cluster
masses $M_{200}$ were inferred through a scaling with richness. This
scaling relation was obtained for a sub-sample of {\small AMASCFI
} candidate clusters having X-ray masses taken from \citet{Gozaliasl}
and \citet{Mirkazemi15}. We refer to \citet{Sarron2018} for details on
this procedure. The typical uncertainty on cluster mass estimate
$M_{200}$ is of order $\sim 0.20-0.25 \ {\rm dex}$.

\subsection{Cosmic filament reconstruction method}
Our reconstruction of the cosmic web from photometric redshifts is
based on the D{\scriptsize IS}crete P{\scriptsize ER}sistent Structure Extractor
\citep[\disperse][]{Sousbie2011}, a software that extracts the cosmic web filaments as ridges of the density field from discrete point distributions
either in 3D or 2D. The extraction is naturally scale-free and robust to noise.\\
\indent The software is based on discrete Morse theory and theory of
persistence. We refer to \citet{Sousbie2011} for a detailed description
of the theoretical grounds of the algorithm as well as its
implementation. Application of the software to astrophysical data can also be found in \citet{Sousbie2011a}. Here we will
briefly present the main features of the algorithm with specific
details of our 3D and 2D use in the relevant sections.

\subsubsection{Cosmic web extraction with \disperse}\label{sec:disperse}
\disperse~computes the density field from the discrete distribution of
points (i.e. galaxy distribution in our application) by computing the Delaunay
tessellation of the points. This is done with the Delaunay Tessellation Field
Estimator \citep[DTFE][]{Schaap2000,Cautun2011} that computes the
density at each galaxy position considering the area (2D) or volume (3D) of the
tessellation cells.\\
\indent Discrete Morse theory then enables the algorithm to find critical
points of the density field, i.e. the points where the gradient of the
field vanishes (minima, saddle points and maxima).\\
\indent The skeleton \citep{Pogosyan2009} is computed as the field lines
joining saddle points to maxima. It is defined as a set of segments
tracing the ridges of the distribution (i.e. the filaments of the
cosmic web).

From there, \disperse~allows to filter only robust structures using as criterium the persistence, defined as the ratio of the density value
at each point of a pair of critical points. In the context of filament
(skeleton) extraction, the pairs of interest are the saddle-maximum
pairs.\\
\indent This ratio (the persistence) quantifies the strength of the pair, i.e. how robust
the topological component due to this pair is to local modification of
the field value. Thus it allows to quantify how significant a
structure is by knowing the noise level in the data. 
In the case of a discrete data set as with a galaxy distribution, \disperse~can
deal with Poisson noise and quantify the robustness of structures in
numbers of $\sigma$.
\begin{figure*}[h!]
  \hspace*{-0.8cm}
  \begin{subfigure}[t]{0.58\textwidth}
    \includegraphics[width=1.\textwidth]{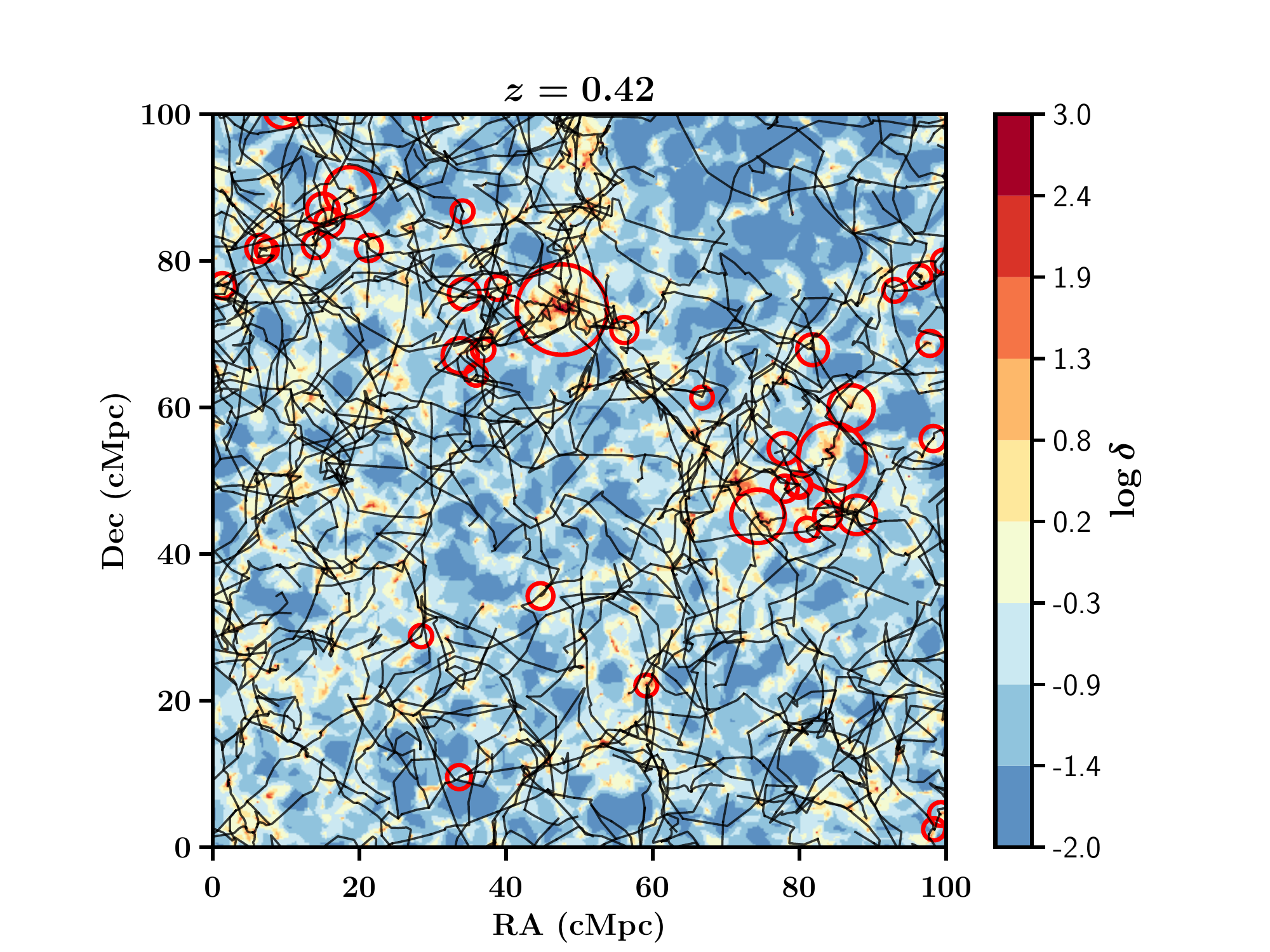}
    \caption{}
    \label{fig:3d}
  \end{subfigure}
  \hspace*{-1cm}
    \begin{subfigure}[t]{0.58\textwidth}
    \includegraphics[width=1.\textwidth]{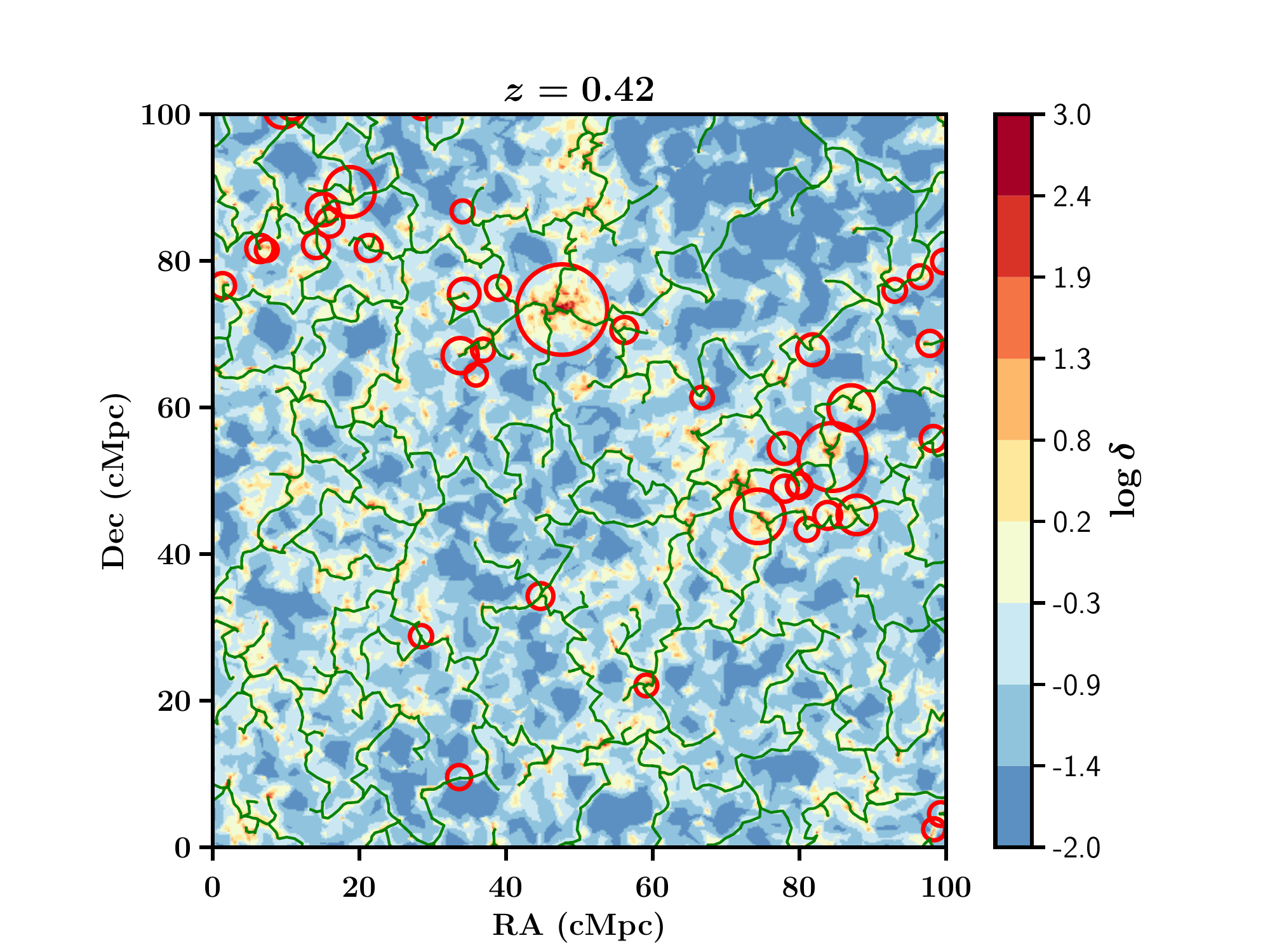}
    \caption{}
    \label{fig:2d}
  \end{subfigure}\\
  \begin{subfigure}[t]{0.55\textwidth}
      \hspace*{0.4\textwidth}
    \includegraphics[width=1.\textwidth]{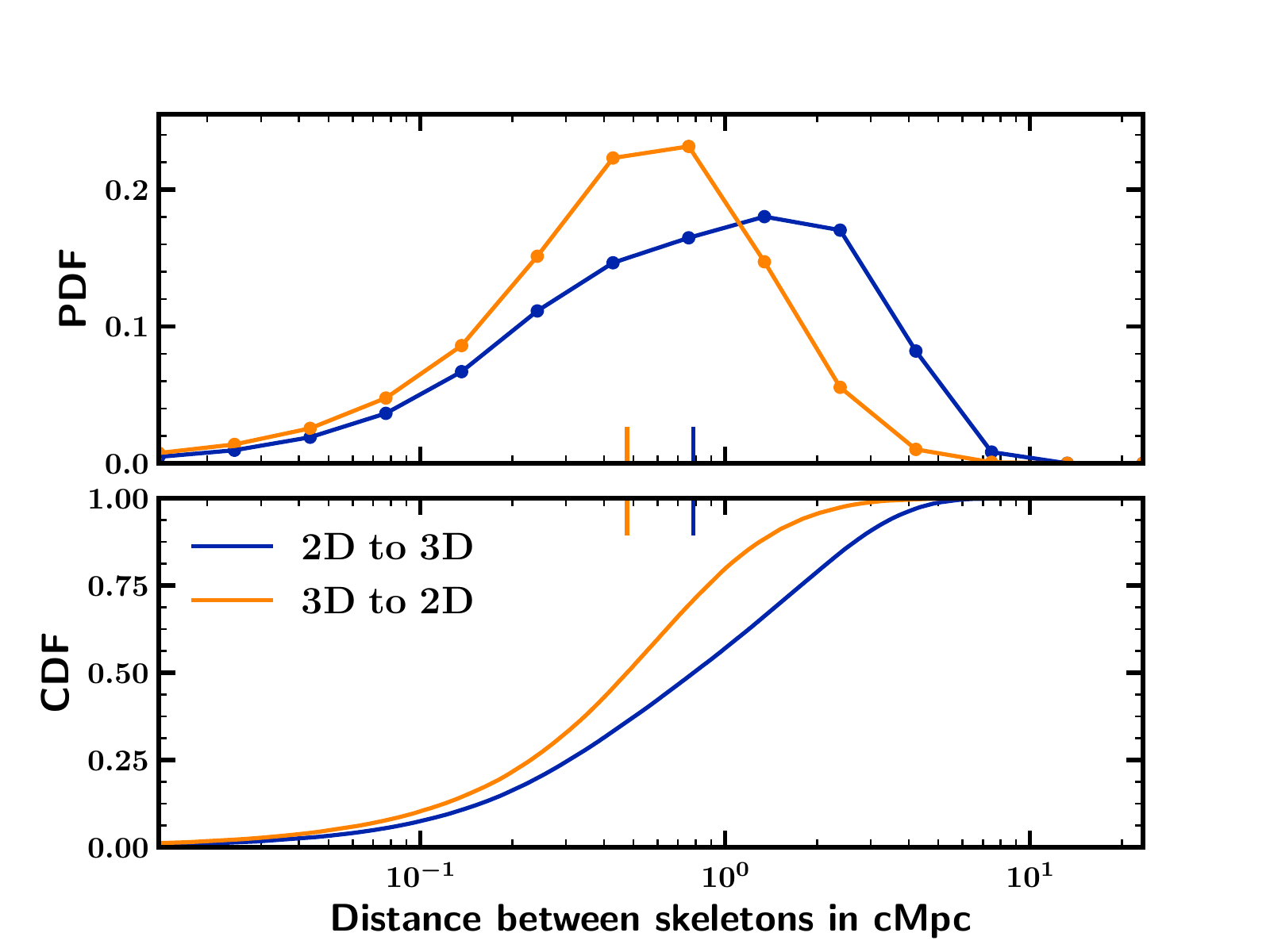}
    \caption{}
    \label{fig:dskel_global}
  \end{subfigure}\\
  \caption{{\sl Top}: $100 \times 100$
    cMpc part of a slice centered at $z = 0.42$ in the mock data. The
    background distribution $\log \delta$ is the logarithm of the DTFE
    obtained from the galaxy distribution in the slice (with
    photo-$z$s). Red
    circles show the positions of halos in the lightcone with redshifts
    $0.41 < z < 0.43$. The radius of the circle is equal to $2 \times
    R_{200}$ for display purposes. In panel (a) black lines are the 3D projected skeleton. In panel (b) green lines are the 2D skeleton. \\
    {\sl Bottom}: Distribution of the distances between skeletons. The distance from the 2D skeleton to the 3D (projected) skeleton is shown in yellow. The distance from the 3D (projected) skeleton to the 2D skeleton is shown in blue. The top panel shows the Probability Distribution Function (PDF) and the bottom panel shows the Cumulative Distribution Function (CDF). The vertical lines and associated error bars on the top panel are the median and error on the median of each distribution.}
  \label{fig:true_vs_zphot}
\end{figure*} 
\subsubsection{3D cosmic web extraction}\label{sec:method_3D}
Here, the 3D skeleton extraction is performed on the
\citet{Merson13} lightcone with \disperse. As
will be detailed in Sect.~\ref{sec:validation}, this is done to assess
the reliability of the 2D reconstruction.

Ideally, one would like to compare the 2D reconstruction to a reference
skeleton obtained from the dark matter (DM) particle distribution in the
lightcone, as galaxies are a biased tracer of the underlying DM
distribution \citep[see e.g.][for a discussion]{Laigle2018}.
Here we chose to work with a lightcone with a large FoV in order
to have meaningful statistics regarding filaments around clusters
of different masses, and to be able to trace filaments on large
scales. The chosen lightcone \citep{Merson13} only allows to access the mock galaxy distribution.
However, since this work does not focus on quantifying the bias of
using galaxies as a tracer of the cosmic web, but rather on the
ability to recover the 3D density field obtained from galaxies with
photometric redshifts, this choice is not penalising. Moreover, reconstruction of the cosmic web from the galaxy distribution in 3D has been shown to trace underlying properties of the density field and its specific
geometry \citep[see e.g.][]{Malavasi2017,Kraljic2018}.

\indent To choose our reference skeleton there are two parameters that we need
to tune. Indeed, as explained in Sect.~\ref{sec:disperse}, the
skeleton is going to change as the detection threshold is
modified. It also depends on the stellar mass limit
of the sample used. Up to $z = 1$ the \citet{Merson13} lightcone is
complete down to $M_* \sim 2.5 \times 10^8 \ {\rm M_\odot}$. 

The aim of this work is to detect and study filaments of galaxies around galaxy
clusters based on photometric redshifts. As detailed in
Sect.~\ref{sec:method_2D}, the photometric redshift uncertainty
prevents us from detecting the leafs and leaflets of the cosmic web. We
thus need to choose both the stellar mass and the significance cut in
3D appropriately, to allow for a meaningful comparison. 

Working only with the most massive galaxies, we might miss some fainter
filaments, but decreasing the mass limit will include more faint
filaments that we will not be able to recover in 2D.
For our reference 3D skeleton, we chose to work with a
stellar mass cut $M_* > 10^{9} \ {\rm M_\odot}$ and a significance threshold set to $5.5 \sigma$. We checked on sub-samples of the mock data that our results are not too sensitive to the exact choice of these parameters.

\subsubsection{2D cosmic web extraction}\label{sec:method_2D}

To identify cosmic filaments in the CFHTLS from photometric redshifts, we apply
the method outlined in \citet{Laigle2018}, with two main modifications.

As in \citet{Laigle2018}, the galaxy catalogue is cut along the redshift dimension in slices of
constant co-moving size. This ensures that the quality of the cosmic
web reconstruction is the same at all redshifts and thus avoids
possible systematics due to increased slice thickness at higher redshifts.

Galaxies belong to the slice corresponding to their photo-$z$. To compute the density field in 2D from these galaxies,
we use the DTFE in two dimensions, each galaxy being weighted by its
probability to be in the slice $p_{\rm gal, slice}$ :
\begin{equation}
\left. p_{\rm gal, \ slice} = \int_{z_{\rm inf}}^{z_{\rm sup}}
P_{\rm gal} (z) \ dz \ \middle / \int_0^\infty P_{\rm gal} (z) \ dz \right.
\end{equation}

\noindent where $z_{\rm inf}, z_{\rm sup}$ are the limits of the
redshift slice.

To deal with boundary conditions, we add a surface of ``guard particles'' outside the bounding box by
interpolating the actual density at the boundary.
Galaxies in masked areas are removed from the catalogue. We do not fill
these regions with fake particles as can sometimes be done
\citep[e.g.][]{Aragon-Calvo2015}.
Once the density is computed we extract the skeleton using \disperse~with a persistence threshold of $2\sigma$ as in \citet{Laigle2018}.
The main modifications of the method in this work are:

$\bullet$ Galaxies in each slice are selected following an absolute magnitude
cut rather than a stellar mass cut as in \citet{Laigle2018}. Here, we select
galaxies with $m_i < m^*_i(z) +1.5$, where $m^*_i(z)$ is computed using
\citet{BC03} single stellar population models calibrated with the
field galaxy luminosity function (GLF) of \citet{Ramos2011}. This cut ensures a good
sampling while limiting the photo-$z$ uncertainty.

$\bullet$ Since our study focuses on cosmic filaments around galaxy
clusters, the centring of the slices is different. Here, the skeleton
around each cluster is reconstructed from a slice centred at the
cluster redshift. This ensures an optimal reconstruction as the bias
from photo-$z$s is minimised.

\section{Validation of the connectivity measurement/filament extraction on mocks}\label{sec:validation}
The aim of this section is to quantify the ability of photo-$z$s to trace accurately the cosmic web in the CFHTLS data. \citet{Laigle2018} showed that high quality photo-$z$s from \citet{Laigle2016} in the
COSMOS survey \citep{Scoville2007} allow to probe the cosmic web influence on galaxies up to high redshift ($z \sim 0.9$). Their sample has a typical photo-z uncertainty of $\sigma_z = 0.008 \times (1 + z)$ at $z < 0.9$.

The photo-$z$ uncertainty in the CFHTLS is $\sim 0.03 \times (1+z)$ (at $m_i < m^*_i(z) +1.5$, $z < 0.7$). As the photo-z uncertainty drives the slice width, in the
CFHTLS slices are chosen to be thicker than in COSMOS. We chose a fixed width of 300 comoving Mpc (cMpc) in the CFHTLS rather than 75 cMpc in COSMOS \citep{Laigle2018}. This choice is justified in Table~\ref{tab:slicewidth}. Considering the increased slice width, we needed to ensure that applying the
\citet{Laigle2018} method to the CFHTLS data allowed to obtain a fair reconstruction of cosmic filaments around clusters.

To do so we worked on our modified CFHTLS-like \citep{Merson13} lightcone to test the quality of the skeleton reconstruction with CFHTLS-like photo-$z$s. Once the slices were chosen, we applied the method described in Sect.~\ref{sec:method_2D} to extract the skeleton at the $2\sigma$ level. We first considered the global skeleton reconstruction in the slice as in \citet{Laigle2018} (Sect.~\ref{sec:global_skel}). We then focused on the reconstruction around clusters studying the connected filaments (Sect.~\ref{sec:connect_skel})

\begin{table}
\centering
\begin{tabular}{cccccc}
\hline
\hline
$z$&$m_i^* + 1.5$ & $\sigma_z$ & $W_{\rm com}$ \\
\hline
\hline
$0.10 < z < 0.20$ &18.40 & 0.040  & 308.6 \\
$0.20 < z < 0.30$ &19.60 & 0.035  & 258.7 \\
$0.30 < z < 0.40$ &20.45 & 0.036  & 254.1 \\
$0.40 < z < 0.50$ &21.05 & 0.039  & 261.8 \\
$0.50 < z < 0.60$ &21.65 & 0.045  & 282.2 \\
$0.60 < z < 0.70$ &22.15 & 0.050  & 299.5 \\
$0.70 < z < 0.80$ &22.65 & 0.072  & 412.0 \\
\hline
\hline
\end{tabular}
\caption{Redshift uncertainties $\sigma_z$ and how they drive the choice of the
  slice thickness. $W_{\rm com}$ is the co-moving width ($\pm
  \sigma_z$) in co-moving Mpc (cMpc) and $m_i^*$ the typical apparent knee magnitude of the field
  Galaxy Luminosity Function (GLF). See text for details.}
\label{tab:slicewidth}
\end{table}

\subsection{Global skeleton}\label{sec:global_skel}

\subsubsection{Distances between skeletons}\label{sec:dskel_global}

To quantify the quality of the photometric reconstruction in the
CFHTLS, we computed the distribution of the distances between the
segments of the 2D skeleton and the projected 3D skeleton \citep{Sousbie2008}.
This is done by computing,
for each segment of the 2D skeleton, the minimum distance to a
segment of the 3D skeleton. This operation can be reversed to compute the distance of the projected 3D skeleton to the 2D skeleton.

Following \citet{Laigle2018}, we define the purity as the proportion of 2D segments that are closer
than $1.5 \ {\rm cMpc}$ from a projected 3D segment. Inversely, the completeness is the proportion of projected 3D segments that are closer than $1.5 \ {\rm cMpc}$ from a 2D segment.

The results are summarised in Table~\ref{tab:dskel2D_to_3D}, where we give the completeness and the purity as well as the median of the distribution of distances. The full distributions of the distances are shown in Fig.~\ref{fig:dskel_global}.

\begin{table*}
\centering
\begin{tabular}{l|cccccc}
\hline
\hline
&Completeness & Purity & median (2D -> 3D) & median (3D -> 2D) \\
\hline
\hline
Global skeleton & 0.70 & 0.91 & $0.48_{-0.32}^{+0.66}$ & $0.79_{-0.58}^{+1.56}$\\
Filaments connected to clusters & 0.71 & 0.66 & $0.73_{-0.63}^{+2.79}$ & $0.41_{-0.34}^{+3.40}$\\
\hline
\hline
\end{tabular}
\caption{Statistics of the distances between the 2D and 3D skeletons
  detected at $2 \sigma$ and $5.5 \sigma$
  respectively. Quoted uncertainties encompass 68\% of the PDF. See text for details.}
\label{tab:dskel2D_to_3D}
\end{table*}

\begin{figure}[t!]
  \centering \includegraphics[width=0.45\textwidth]{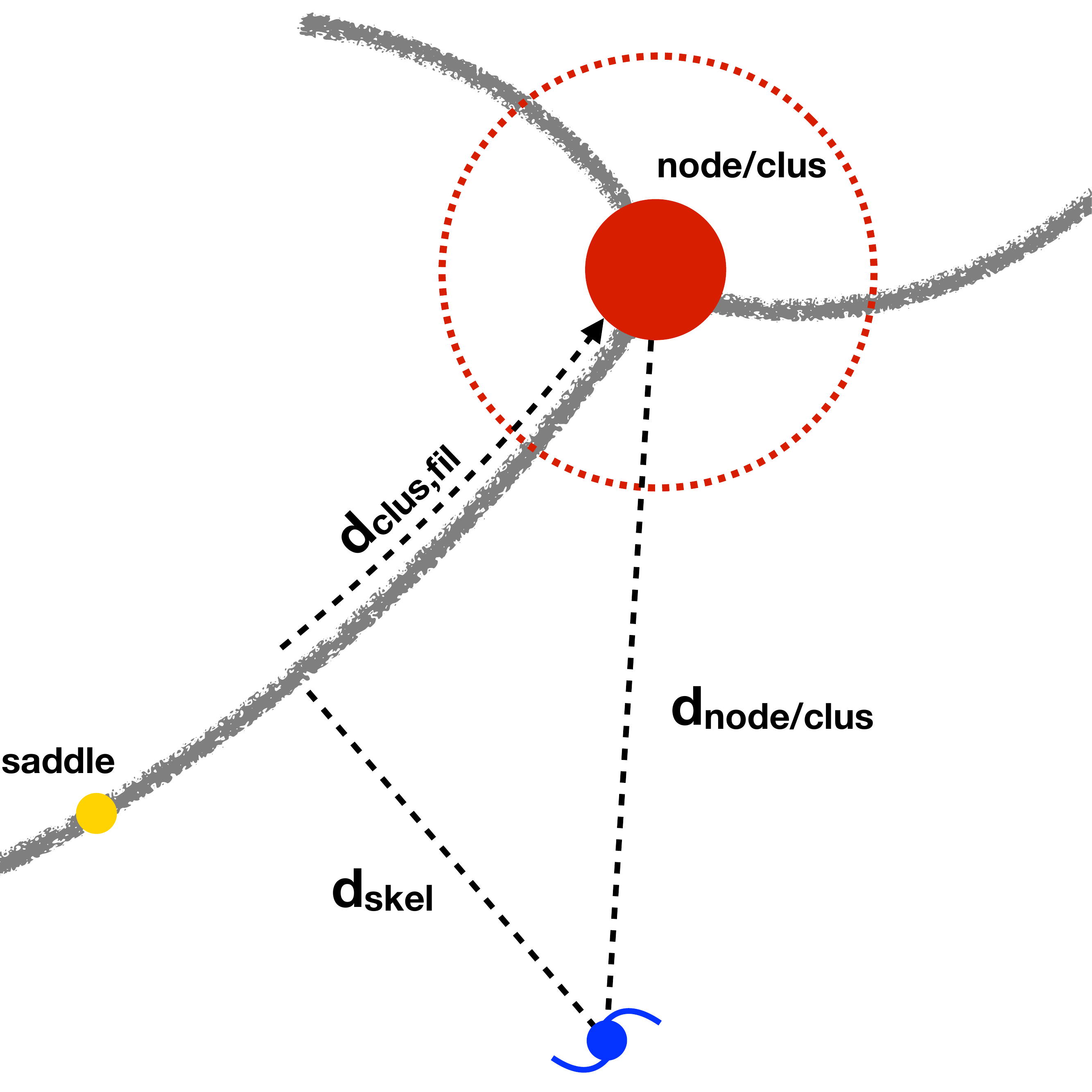}
  \caption{Schematic view of the cosmic web as traced by \disperse~that defines the different distances used in this chapter. The
    grey lines are the filaments. The red and yellow points are
    respectively a node and saddle point. The red dashed circle shows the
    exclusion radius around the node. $d_{\rm skel}$ is the distance
    to the skeleton, $d_{\rm node/clus}$ the distance to the node and
    $d_{\rm clus,fil}$ the distance to the cluster along the filament.}
  \label{fig:dist_def}
\end{figure}

We note that with the thresholds chosen in 3D ($5.5 \sigma$) and in 2D ($2 \sigma$), when the 3D skeleton is projected in the redshift slices, there are about twice as many 3D projected filaments than 2D filaments. This should impact our selection function by biasing results towards higher purity and lower completeness compared with the case where $N_{\rm fil,2D} \equiv N_{\rm fil,3D}$. Yet, when choosing a higher significance threshold in 3D, from visual inspection, it was clear that some filaments having a counterpart at $5.5\sigma$ were
considered as false detections, thus biasing our purity towards lower values.

\subsubsection{Stellar mass gradients towards filaments}
We have shown that the 2D skeleton is a good tracer of the projected 3D structures. However, to use the 2D skeleton in real data to
infer filament properties we need to confirm that we can trace signals that actually exist in 3D from our reconstruction.
Here, we consider the stellar mass gradient towards filaments
observed in 3D by \citet{Malavasi2017} and \citet{Kraljic2018}. \citet{Laigle2018} showed that this
gradient is recovered by the 2D reconstruction given the COSMOS precision. Here, we study whether it is recovered  in our CFHTLS-like mock or not.

As in \citet{Laigle2018} and \citet{Kraljic2018} we remove the contribution of nodes by
removing galaxies too close to nodes, in order to compute the
effect of filaments alone. Indeed, there is a known stellar
mass gradient towards nodes (i.e. clusters and groups) that we do not want to account for here. Therefore, we stress that the stellar mass gradients we detect are valid at a given scale.\\ 
\indent Stellar mass gradients are measured in three stellar mass bins: $9.5 < \log{M_* / M_\odot} \le 10$, $10 < \log{M_* / M_\odot} \le
11$ and $ 11 < \log{M_* / M_\odot} \le 12$. In each bin, we measured the distance of all galaxies to their closest filaments as illustrated in Fig.~\ref{fig:dist_def}.

\begin{figure*}[t!]
  \hspace*{-0.9cm} 
  \begin{subfigure}[t]{0.5\textwidth}
    \includegraphics[width=1.15\textwidth]{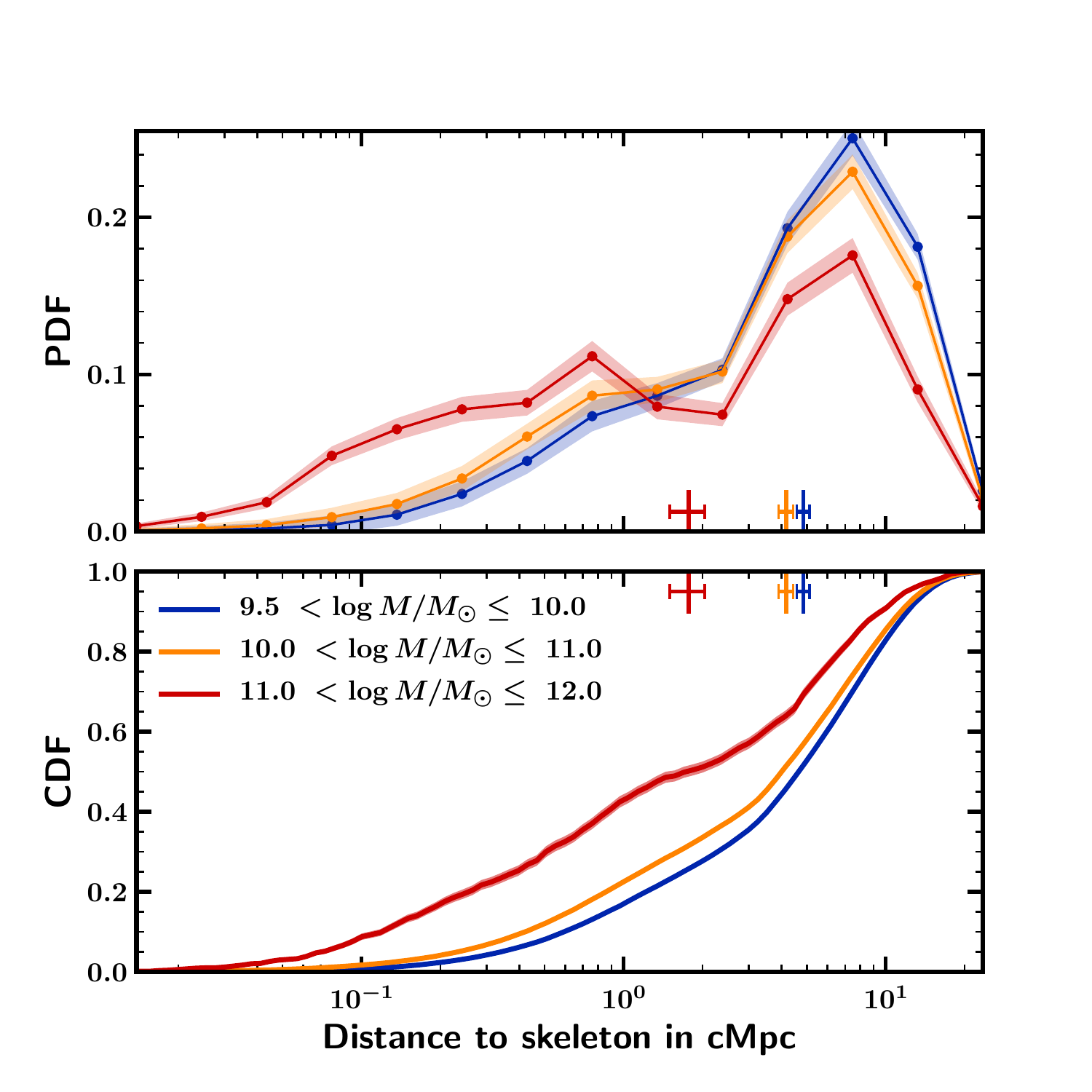}
    \caption{3D at $5.5\sigma$}
    \label{fig:3D_grad}
  \end{subfigure}
  ~~~~
  \begin{subfigure}[t]{0.5\textwidth} \includegraphics[width=1.15\textwidth]{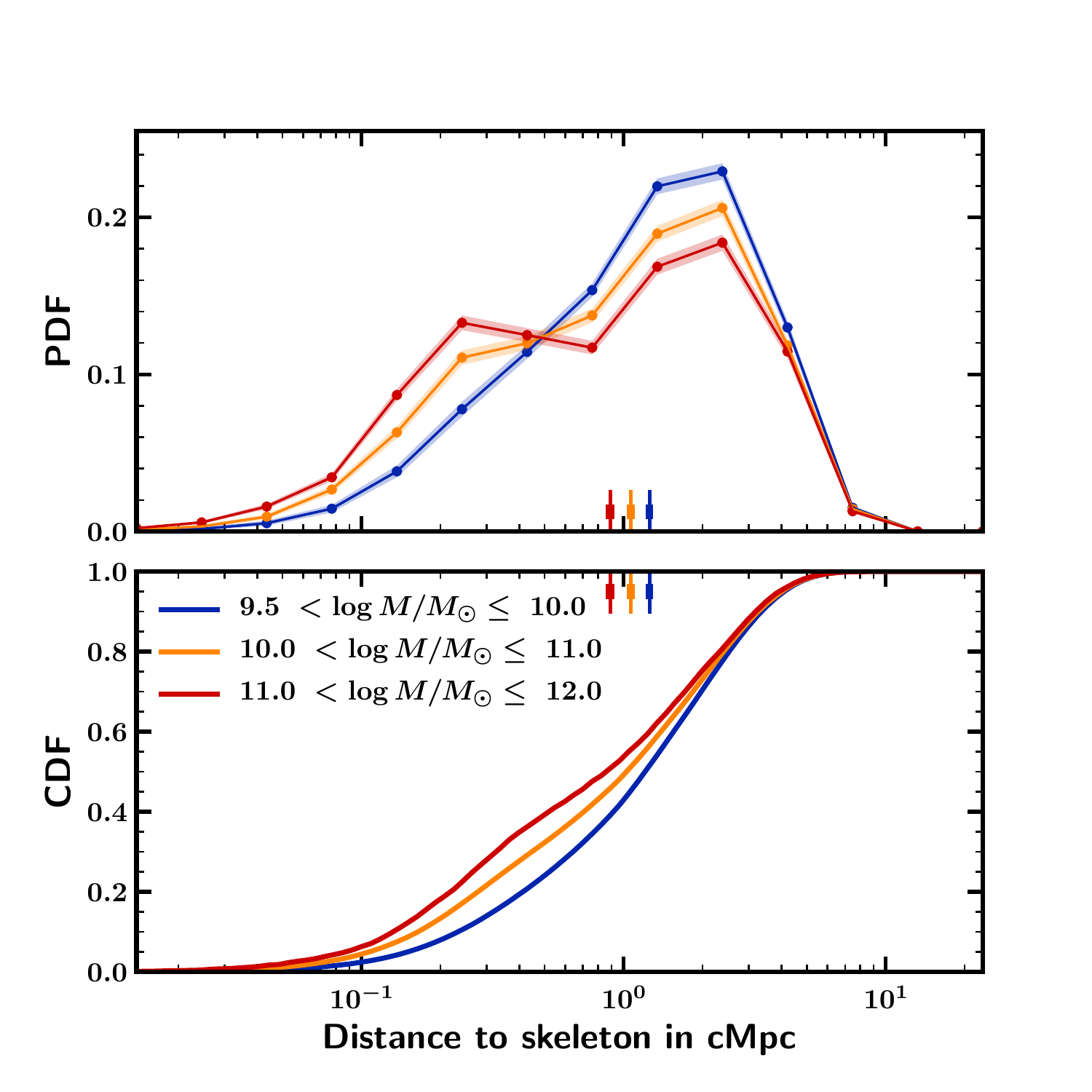}
    \caption{$\sigma_0 = 0.03$}
    \label{fig:sig0.03_grad}
  \end{subfigure}
  \caption{Stellar mass gradients towards filaments detected in 3D
    and in 2D using the mock data. We split the galaxy catalogue in three
    mass bins for which we compute the distribution of the galaxy
    distances to the skeleton after having removed galaxies close to
    nodes (see text for details). Red is for $11 < \log M_* / M_\odot
    < 12$, yellow for $10 < \log M_* / M_\odot
    < 11$ and blue for $9.5 < \log M_* / M_\odot
    < 10$. In each subfigure, the top distribution is the PDF of the
    distances to the skeleton. The filled areas around the
    curves represent the 68\% confidence limits computed from 10000
    bootstrap re-samplings of the distributions. The bottom panels show
    the CDF. The vertical lines and
  associated error bars are the median and error on the median of each
  distribution.}
  \label{fig:SM_grad}
\end{figure*}

In 3D, we remove all galaxies that are closer than 3.5 cMpc from a node. This is a rather conservative choice. Indeed, this value is larger than $\sim 2 R_{200}$ for the most massive halo in the lightcone.
We take the $5.5\sigma$ skeleton as our reference skeleton for the 3D measurement. We tested that when lowering the significance
threshold the distances of galaxies to filaments increase as fainter groups are classified as nodes and thus removed. The inverse is true when
increasing the detection  threshold, as the less massive halos are not detected any more, so their galaxies are included in the distance to filament statistics.The results are shown in Fig.~\ref{fig:SM_grad}. A significant stellar mass gradient towards filaments is detected in 3D, with more massive galaxies
lying closer to filaments.\\
\indent In 2D, we adopt a projected exclusion radius of 1 cMpc. This is less
conservative. Yet, adopting a larger value would drastically reduce the available statistics, while a smaller radius increases the influence of the nodes in the measurement, making the 1 cMpc choice a good compromise. The results  are presented in Fig.~\ref{fig:SM_grad}. The stellar mass gradient towards
filaments can still be observed in 2D even though it is dimmed
compared to the 3D signal.\\
\begin{table}
\centering
\begin{tabular}{c|cccc}
\hline
\hline
 \multicolumn{3}{c}{Median values of the PDF (in cMpc)}\\
  $\log M_{*}$ & 3D & 2D mock data\\
\hline
  \hline
$9.5 < \log{M_{*} / M_\odot} < 10$&  $4.37 \pm 0.02$ & $1.26 \pm 0.01$\\ 
$10 < \log{M_{*} / M_\odot} < 11$& $3.77 \pm 0.02$  & $1.09 \pm 0.01$\\ 
$ 11 < \log{M_{*} / M_\odot} < 12$& $1.51 \pm 0.22$ &  $0.88 \pm 0.03$\\ 
\hline
\hline
\end{tabular}
\caption{Median values of the galaxy distances to the skeleton in three mass
bins, in the original 3D lightcone and in the three toy-model mocks
considered in this work. Error bars are the error on the median computed from 100
bootstrap re-samplings of the original distribution. Values are in cMpc.}
\label{tab:SM_grad}
\end{table}
\indent The median values of the PDF for the different mass bins and different
cases are reported in Table~\ref{tab:SM_grad}.

\subsection{Reconstruction around clusters}\label{sec:connect_skel}

\subsubsection{Cluster connectivity}\label{sec:connectSIM}
The quality of the reconstruction in the infall regions of clusters can
also be assessed by studying the cosmic connectivity
$\kappa$ of clusters at the nodes of the cosmic web, i.e. the number of
cosmic filaments connected to a cluster.

Cosmic connectivity is expected to scale with cluster mass, more
massive clusters being more connected
\citep[see e.g.][]{Aragon-Calvo2010,Gouin2017,Codis2018,Fordinprep}, though with
large intrinsic scatter \citep[see in particular][]{Aragon-Calvo2010}. Finding  such a
correlation from our 2D filaments would give an
independent confirmation of our skeleton reconstruction quality,
particularly in cluster infall regions.

In the lightcone, we computed the connectivity considering the 3D
skeleton. The same measurement was carried out in our CFHTLS-like mock
data using the 2D reconstruction.

In the skeleton extracted with \disperse, all nodes
are linked to one or several saddle-points through filaments such that one node may be connected to
several filaments.
Formally, in the skeleton extracted by \disperse, two filaments leading
to the same node can become infinitely close but still be counted
separately, as they are both topologically robust. However, they represent
only a single filament physically speaking. Thus to avoid double
counting, these are merged into a single filament and a bifurcation
point is added where they diverge.
Our measure of a node's connectivity is then the number of physical
filaments departing from the node and crossing the sphere (3D) or
circle (2D) of radius $1.5$ cMpc centred on the node. This definition
is slightly different from that of \citet{Fordinprep} where they took 
a radius of $1.5 \times R_{200}$. This is because our CFHTLS candidate
cluster mass estimate has a high uncertainty ($ \sim 0.20-0.25 \
{\rm dex}$) compared to theirs. Using the value of $R_{200}$ computed from our estimated $M_{200}$ would thus introduce noise in our connectivity measurement.

To compute the number of filaments connected to a given cluster, we first match
the cluster to the nodes detected by \disperse~and choose the node
which is closest to the cluster. If no node is found at a
distance smaller than the cluster $R_{200}$, the cluster is marked as
unmatched and not considered in the analysis. Here $R_{200}$ is computed from
the halo mass $M_{200}$ following:

\begin{equation}
  R_{200} = \left ( \frac{3 \, M_{200}}{4 \pi \, 200 \, \rho_{c}(z)} \right )^{\frac{1}{3}}
\end{equation}\label{eq:r200}

We then define the connectivity of a given cluster as the connectivity
of the node it was matched with.

\begin{figure}[h!]
  \centering
  \includegraphics[width=0.5\textwidth]{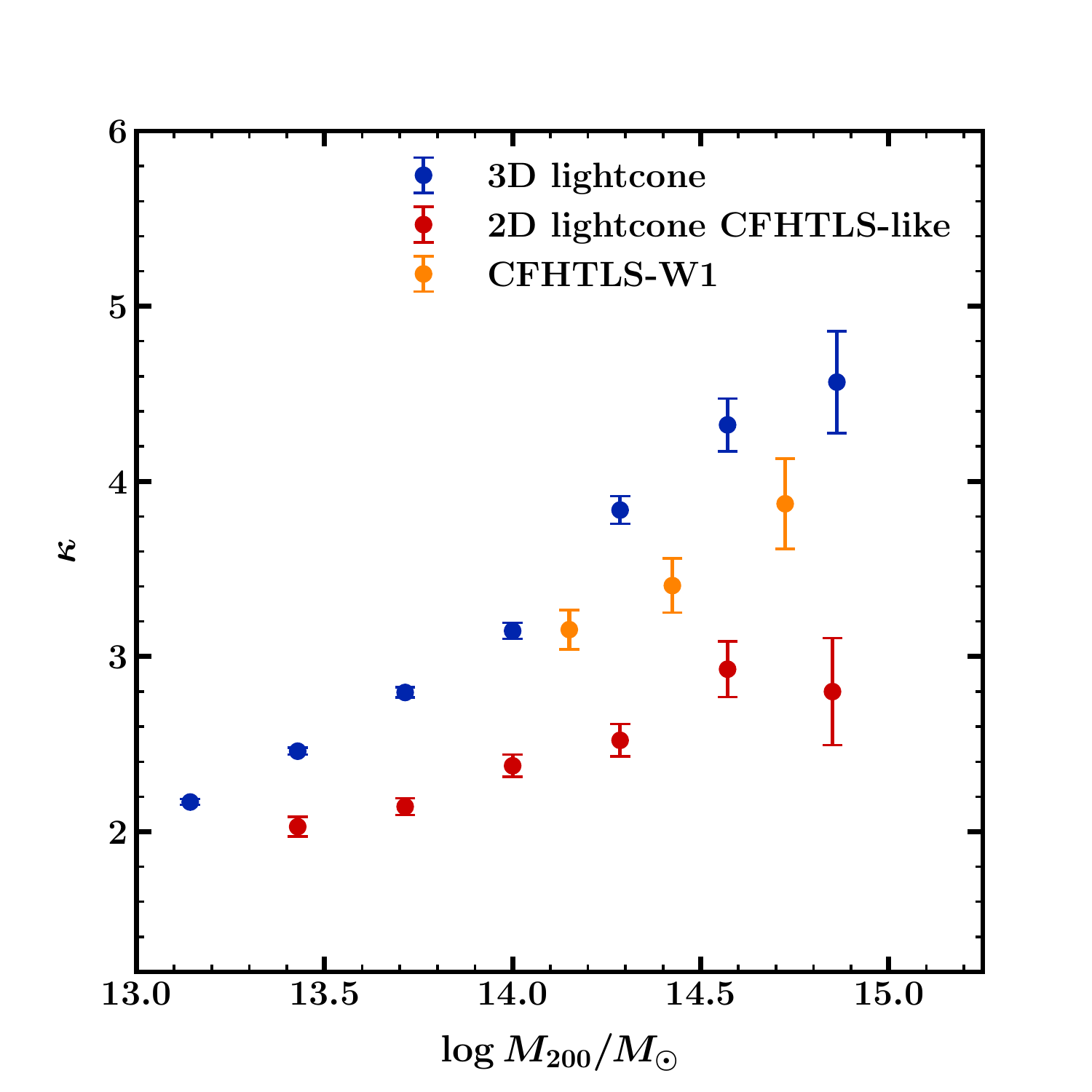}
  \caption{Mean connectivity $\kappa$ in bins of halo mass $\log M_{200}$ in 3D
  in the \citet{Merson13} lightcone (blue), and 2D in the CFHTLS-like mock (red) and in the CFHTLS (yellow). Error bars are standard errors on the mean.}
  \label{fig:connectSIM}
\end{figure}

\begin{figure*}[h!]
  \hspace*{-0.8cm}
  \begin{subfigure}[t]{0.55\textwidth}
    \includegraphics[width=1.\textwidth]{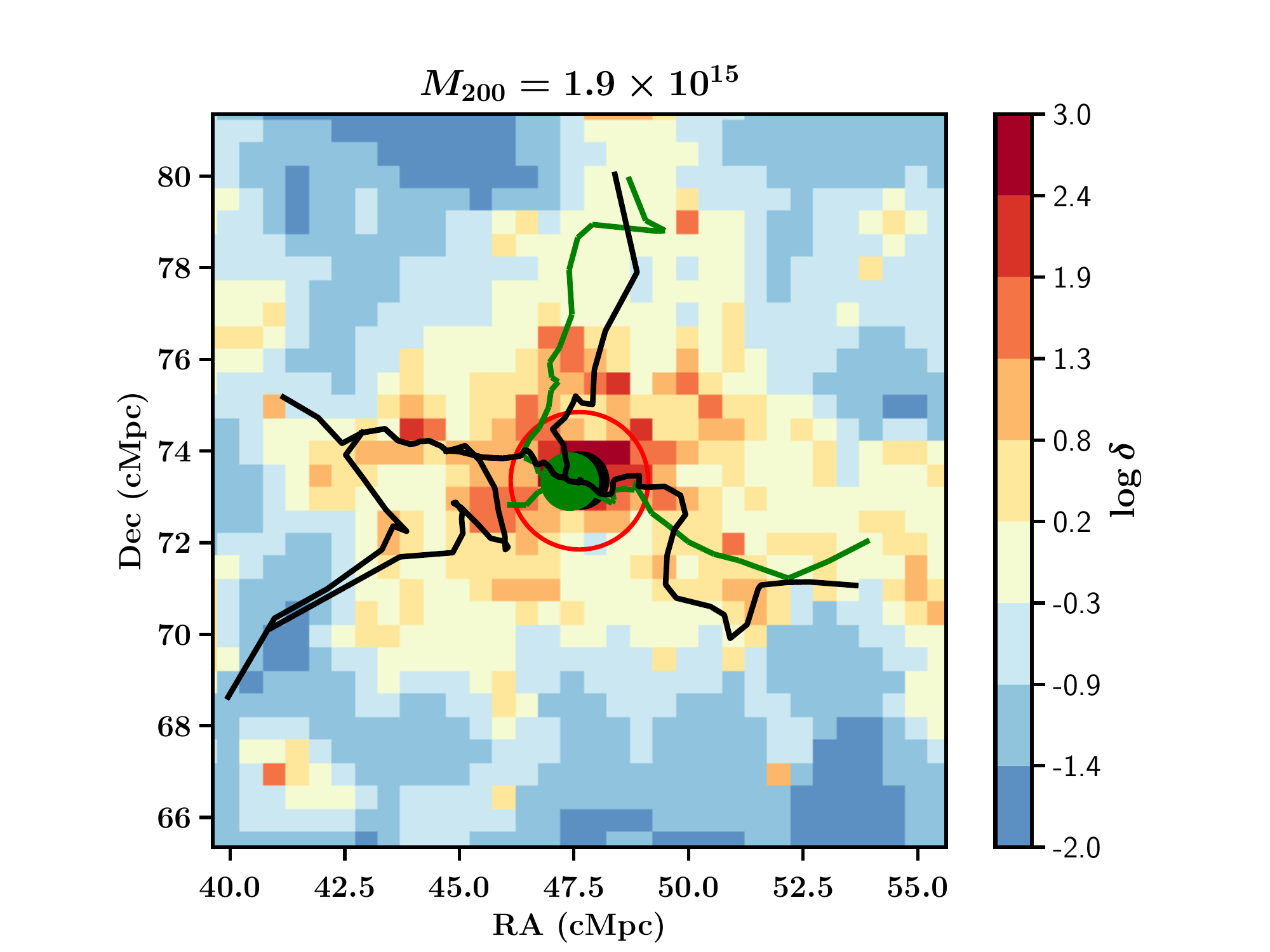}
  \end{subfigure}
  \hspace*{-0.8cm}
  \begin{subfigure}[t]{0.55\textwidth}
    \includegraphics[width=1.\textwidth]{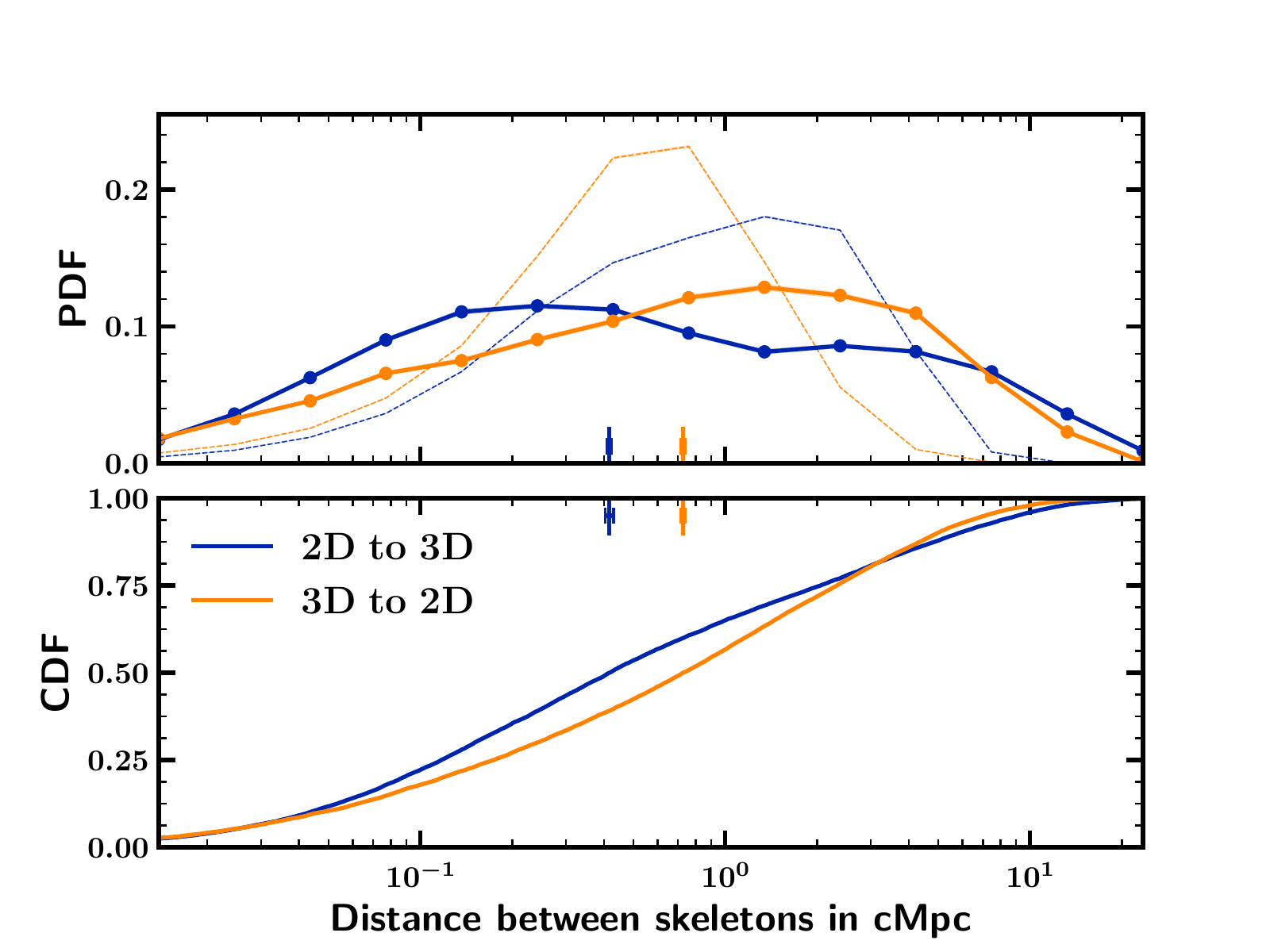}
  \end{subfigure}\\
  \caption{{\sl Left}: Zoom on the most massive halo in the mock. The red circle shows the radius used to compute the
  connectivity ($1.5$ cMpc). We show only the filaments of
  the skeleton connected to the cluster according to the 3D
  reconstruction (black) and 2D reconstruction (green).The
    background distribution $\log \delta$ is the logarithm of the DTFE   obtained from the galaxy distribution in the slice (with photo-$z$s). {\sl Right}: Distribution of the distances between skeletons. Solid lines show the distribution for the filaments connected to clusters ($M_{200} > 10^{14} M_\odot$). The distribution for the global skeleton is shown in the background as thin transparent lines. The top panel shows the PDF and the bottom panel shows the CDF. The vertical lines and associated error bars on the top panel are the median and error on the median of each distribution.}
  \label{fig:dskel_clus}
\end{figure*} 

The expected increase of connectivity with cluster mass is recovered as seen in Fig.~\ref{fig:connectSIM}. Error bars are the standard error on the mean. The standard
deviation of the distribution in each bin is actually much larger, due
to a large intrinsic scatter in the $\kappa - \log M_{200}$ scaling
relation.
We performed a Spearman correlation test to formally check for correlation between connectivity and mass. In each case, we found a weak but strongly significant correlation. Results are reported in Table~\ref{tab:spearman_k}.
\begin{table}
\centering
\begin{tabular}{c|cccc}
\hline
\hline
   & $r_{s}$ & $p-$value\\
\hline
\hline
3D mock data &  $0.29$ & $<< 10^{-15}$\\ 
2D CFHTLS-like mock data&  $0.14$ & $< 10^{-11}$\\ 
CFHTLS-W1 & $0.17$ &  $<10^{-4}$\\ 
\hline
\hline
\end{tabular}
\caption{Results of the Spearman correlation test for $\kappa$ vs $M_{200}$ in the three cases studied : 3D mock data, 2D CFHTLS-like mock data and CFHTLS-W1 respectively.}
\label{tab:spearman_k}
\end{table}

This comparison highlights that the 2D connectivity is biased towards lower values compared to the 3D connectivity. This could come from the fact that some of the 3D
filaments are along the line of sight and cannot be recovered in
2D. This could also be due to faint filaments getting blurred into the
noise (due either to the slice thickness or to the photo-$z$ uncertainty).
Moreover, this test seems to indicate that the bias is slightly
dependent on halo mass, the slopes of the 2D and 3D scaling relations
being different.

We note that in both cases we did not find any redshift evolution of the connectivity in the range $0.1 < z < 0.7$. This is compatible with measures by \citet{Codis2018}, where only a weak evolution is found
between $z=0$ and $z=1.3$.\\

\subsubsection{Distance between skeletons in cluster infall regions}\label{sec:dskel_clus}


To go beyond connectivity and check if the 2D cosmic filaments connected to clusters are representative of the 3D connected cosmic filaments, we computed the distances between these filaments in the mocks. This allows to check if their directions statistically agree, which could not be assessed from the connectivity measurement.

The extraction of connected filaments is done as in Sect.~\ref{sec:connectSIM}. The distances between the 3D and 2D connected filaments are computed as in Sect.~\ref{sec:dskel_global}. Results are displayed in Fig.~\ref{fig:dskel_clus}.

The median of distances between 2D and 3D filaments is higher with this method than when comparing the global skeletons (see Fig.~\ref{fig:dskel_global}). 
This is due to projection effects existing in the global method, as some filaments that may be in the background or in the foreground and thus not connected to the clusters are included in the global reconstruction. Our results are not affected by these projections here. 
On the other hand, the median of distances between 3D and 2D filaments is lower, as fewer 3D filaments have no counterpart in 2D.

This reflects in the purity of the reconstructed filaments (computed as in Sect.~\ref{sec:dskel_global}, matching segments closer than 1.5 cMpc). Indeed this drops to 66\%, meaning that around two thirds of the reconstructed connected filaments correspond to actual 3D filaments feeding the clusters. The analysis of Sect.~\ref{sec:dskel_global} then indicates that about 20\% of reconstructed filaments correspond to 3D filaments appearing in the slice because of projection effects, and 10\% are just false detections due to projection effects in the galaxy distribution and photometric redshift errors. On the other hand, the reconstruction allows to detect $\sim 70 \%$ of true 3D connected filaments.

\subsection{Caveats and limitations of the method}\label{sec:caveats}
One major limitation of our reconstruction method is obviously the fact that, as
we work in 2D we are sensitive to projection effects. If \disperse~deals with Poisson noise and should thus clean properly spurious
alignments, our results may be affected by coherent projection effects
due to walls or filaments oriented in the direction of the
slicing. The former might then be detected as a filament and the
latter as a node.

The amplitude of this effect is difficult to quantify, but it might
play a role when computing the stellar mass gradient towards filaments
in 2D. Indeed, it can be seen in Fig.~\ref{fig:SM_grad} that the behaviours in 3D and 2D are not the
same. Massive galaxies in the 3D skeleton present a pronounced
gradient that does not appear as clear in 2D. This may be due to the
fact that, as \citet{Kraljic2018} showed, in addition to the
stellar mass gradient towards filaments, there is also a stellar mass
gradient towards walls as well, that we pick up in our 2D analysis.

Despite this limitation, we reach a quite high purity in our 2D filament detection at the CFHTLS accuracy ($\sim 90\%$ in the global skeleton reconstruction and $\sim 70\%$ for filaments connected to clusters). We can thus use our filament reconstruction on the CFHTLS T0007 data to study how filaments impact galaxy properties.
\section{Cosmic filaments around AMASCFI clusters in the CFHTLS}\label{sec:CFHTLS_results}
\begin{figure*}[h!]
  \begin{subfigure}[t]{0.52\textwidth} \includegraphics[width=1.\textwidth]{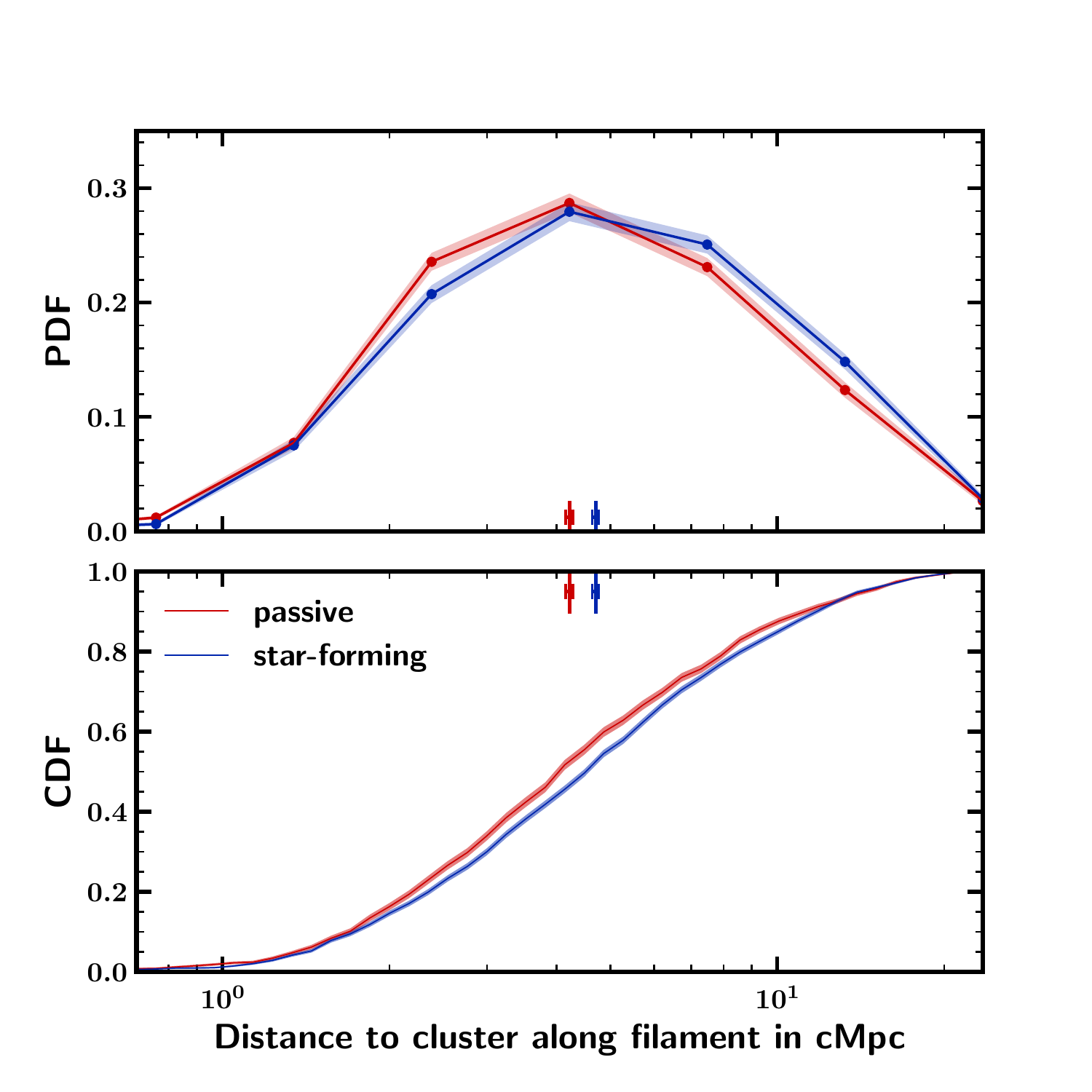}
    \caption{$0.15 < z \le 0.4$}
    \label{fig:dgal_to_clus_early_Mall0}
  \end{subfigure}
  \begin{subfigure}[t]{0.52\textwidth}     \includegraphics[width=1.\textwidth]{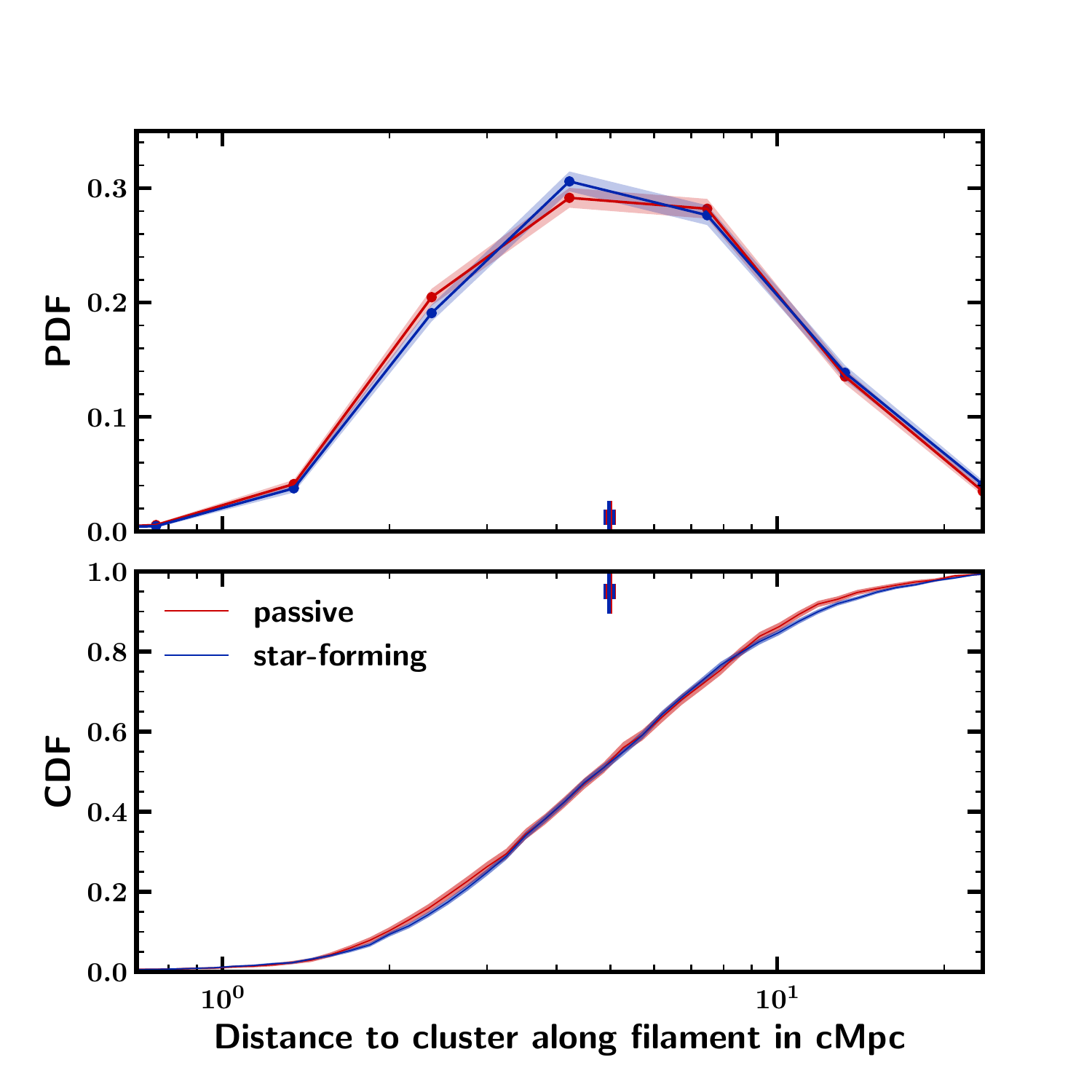}
    \caption{$0.4 < z < 0.7$}
    \label{fig:dgal_to_clus_late_Mall0}
  \end{subfigure}
  \caption{Distances to {\small AMASCFI} clusters along filaments of
    passive galaxies (red) and star-forming galaxies (blue) respectively at low (a) and high (b)
    redshifts. The top distribution is the PDF of the
    distances to the skeleton. The filled areas around the
    curves represent the 68\% confidence limits computed from 100
    bootstrap re-samplings of the distribution. The significance of
    the difference between the passive and star-forming galaxy
    distributions is written at each sampling point. The vertical lines and
  associated error bars are the median and error on the median of each distribution.\\
    The bottom panel shows
    the CDF. }
  \label{fig:dclus_z_type}
\end{figure*}
We then proceeded to study the properties of cosmic filaments around AMASCFI clusters in the CFHTLS T0007 data. We particularly focused on the role filaments may play in environmental quenching by pre-processing galaxies infalling in galaxy clusters, by comparing the properties of passive and star-forming galaxies. To this aim, we reconstructed the filaments using the technique described in Sect.~\ref{sec:method_2D}. The ability of this technique to statistically trace the true 3D cosmic web at the CFHTLS precision was demonstrated in Sect.~\ref{sec:validation}.


\begin{figure*}
  \begin{subfigure}[t]{0.52\textwidth}  \includegraphics[width=1.\textwidth]{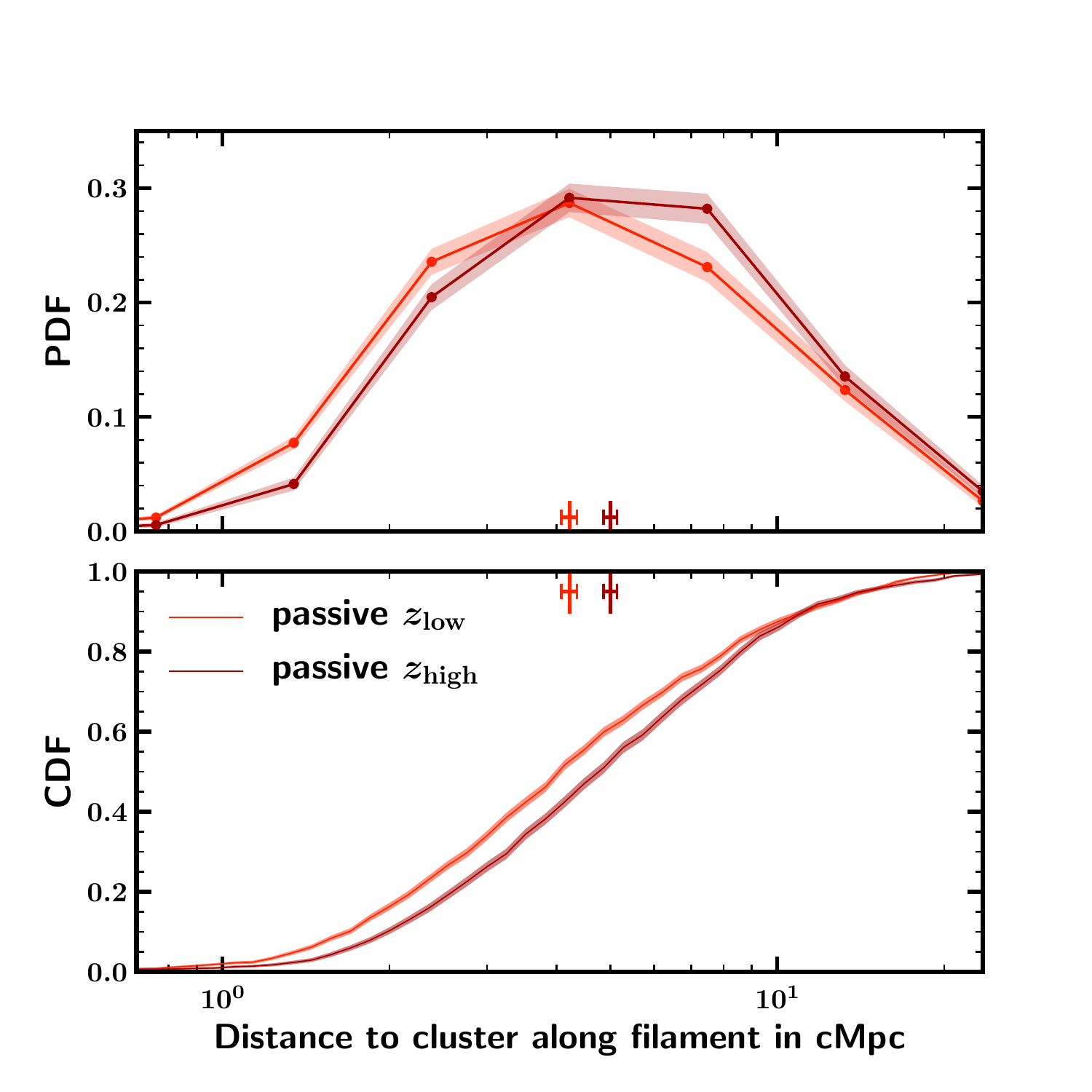}
    \caption{passive galaxies}
    \label{fig:dgal_to_clus_early_Mall}
  \end{subfigure}
  \begin{subfigure}[t]{0.52\textwidth} \includegraphics[width=1.\textwidth]{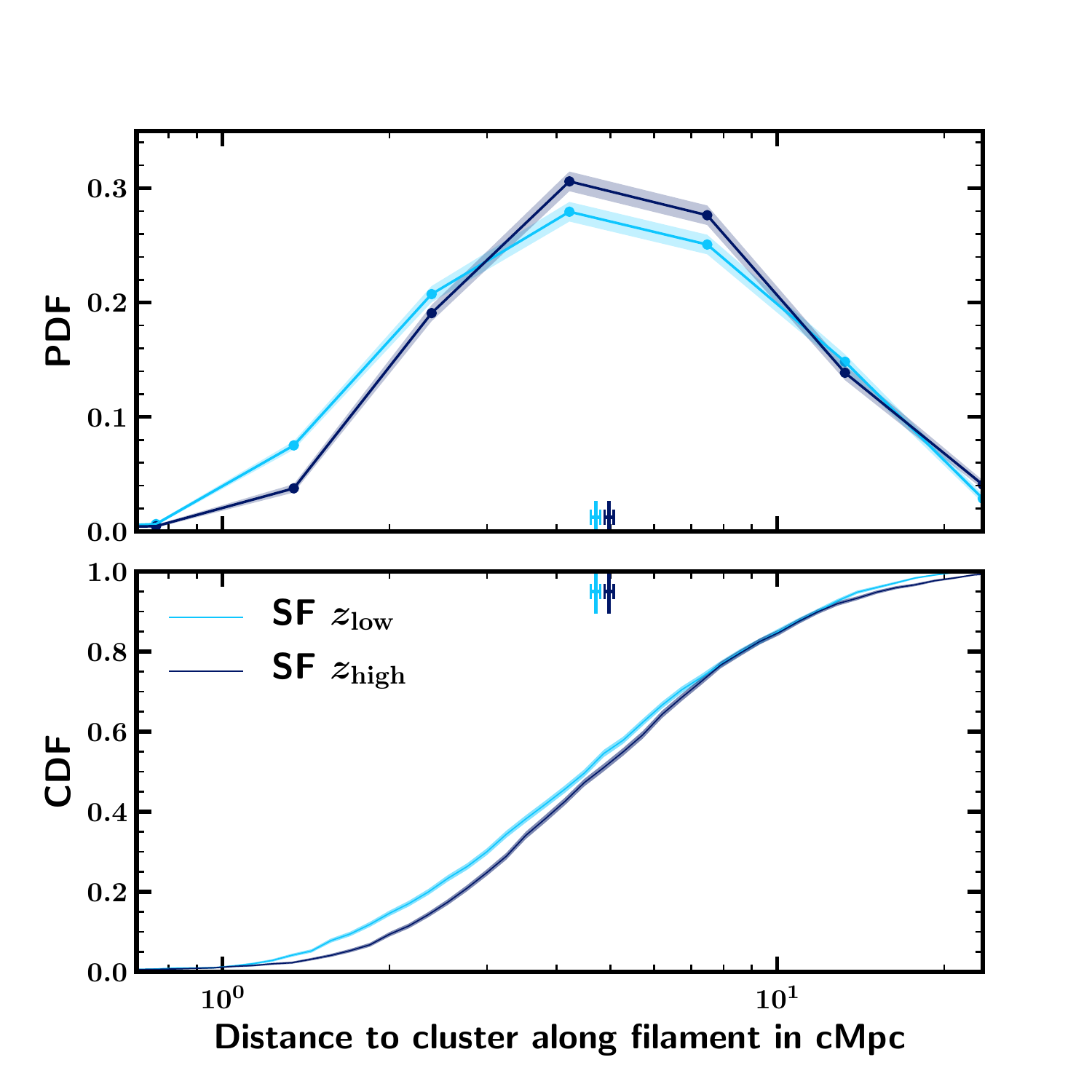}
    \caption{star-forming galaxies}
    \label{fig:dgal_to_clus_late_Mall}
  \end{subfigure}
  \caption{Distances to {\small AMASCFI} clusters along filaments of
    respectively passive galaxies (a) and star-forming galaxies (b) at high (dark) and low (light)
    redshift. See Fig.~\ref{fig:dclus_z_type} for details on symbols.}
  \label{fig:dclus_type_z}
\end{figure*}

\begin{table*}
\centering
\def\arraystretch{1.05}
\begin{tabular}{c|cc|c}
\hline
\hline
 \multicolumn{3}{r}{Median values of the PDF (in cMpc)}\\
 & passive galaxies & star-forming galaxies & $\Delta_{{\rm passive}-{\rm star-forming}}$\\
\hline
  \hline
$0.15 < z_{clus} < 0.70$ & $4.50 \pm 0.05$ & $4.98 \pm 0.05$ & $7.6\sigma$\\
  \hline
$0.15 < z_{clus} \le 0.40$ & $4.22 \pm 0.08$ & $4.71 \pm 0.07$ & $4.9\sigma$\\ 
$0.40 < z_{clus} < 0.70$ &  $5.00 \pm 0.13$ & $4.98 \pm 0.10$ & NS\\ 
& $----$ & $----$ & \\
$\Delta_{z_{\rm low}-z_{\rm high}}$ & $5.0\sigma$ & $2.3\sigma$ & \\
\hline
\hline
\end{tabular}
\caption{Median values of the PDF of the filament galaxy distances to AMASCFI clusters
  along filaments for passive galaxies and star-forming galaxies in different redshift bins. NS stands for 'not significant' and $\Delta_{X-Y}$ is defined in eq.~\ref{eq:Delta}}
\label{tab:medians_type_z}
\end{table*}


\subsection{Galaxy cluster connectivity}\label{sec:connect}

To compute the connectivity of {\small AMASCFI} clusters, we used the
same method as in Sect.~\ref{sec:connectSIM}. The difference is that
here, we do not know the exact position of the cluster but rather its
estimate as computed by {\small AMASCFI}. So this should introduce some bias in our connectivity measurement. The same is true for the
mass. The cluster $R_{200}$ is computed from the {\small AMASCFI} mass
estimate following eq.~\ref{eq:r200}.

We investigated the scaling between the connectivity and cluster mass
in the CFHTLS-W1 field from {\small AMASCFI} clusters using three mass bins
$14 < \log M_{200} / M_\odot \le 14.3$, $14.3 < \log M_{200} / M_\odot
\le 14.6$, and $\log M_{200} / M_\odot > 14.6$, as in
\citet{Sarron2018}.

We do recover the expected connectivity increase with cluster
mass as can be seen in Fig.~\ref{fig:connectSIM} (yellow points). Such a scaling was also
found in lower mass groups in the COSMOS survey by
\citet{Fordinprep}. In this
figure, the error bars are standard errors on the mean. The standard
deviation of the distribution in each bin is actually much larger, due
to a large intrinsic scatter in the $\kappa - \log M_{200}$ scaling
relation.
We note that the mean connectivity at a given mass is higher in the
CFHTLS data when compared to the value obtained in the lightcone. This could be due to the difference in number counts between the mock and the data mentioned in Sect.~\ref{sec:data_method}. 

\subsection{Galaxy-type gradients towards clusters}\label{sec:dclus_CFHTLS}

Since galaxies fall along filaments onto clusters, we might
expect to see a redshift dependence of their median distance to
clusters along filaments. If galaxies are quenched inside filaments in
their fall towards clusters, we then expect to see a colour-type gradient toward
clusters inside filaments.\\
\indent \disperse~allows one to carry such a measurement, as for each cluster
the connecting filaments are well defined (see
Sect.~\ref{sec:connectSIM}). Moreover we showed in Sect.~\ref{sec:dskel_clus} that the 2D reconstructed filaments are a good tracer of the actual 3D filaments feeding clusters ($\sim 70\%$ completeness and purity). 
We can thus compute the distribution of the distances to {\small AMASCFI} clusters
along their connecting filaments.\\
\indent Galaxies at a distance $d_{\rm skel} < 1 \ {\rm cMpc}$ are considered
filament members. This definition is coherent with that of
\citet{Tempel2014} (who took a radius of $0.5  h^{-1}$ (physical) Mpc at $z <
0.15$, which corresponds to $1$ cMpc at $z = 0.4$) and more restrictive than \citet{Martinez2016} who
used a radius of $1.5 h^{-1}$ (physical) Mpc at $z < 0.15$. 

This definition of the comoving radial size is fixed for all filaments connected to the {\small AMASCFI} clusters. We plotted the radial profile (without background substraction) of these filaments as a function of several parameters. In particular, we checked the dependency of the radial profile as a function of the cluster redshift, cluster mass $M_{200}$, and distance to the cluster along the filament.
We found no redshift evolution nor dependence on cluster mass (no significant difference between the distributions). The radial profiles show no significant variation at $d_{\rm clus,fil} > 1.5$ cMpc. At $d_{\rm clus,fil} \le 1.5$ cMpc, filaments become significantly more concentrated, showing that the influence of the cluster on the filaments becomes significant at this distance only - corresponding roughly to the virial radius.

We computed the distance to the connected
cluster along the filament axis $d_{\rm clus,fil}$ for these galaxies. We did not account
for  $d_{\rm skel}$ in the measurement (see Fig.~\ref{fig:dist_def}
for the distance definitions).
Since we are interested in the role played by cosmic filaments in quenching, we
remove all galaxies at a distance $d_{\rm clus} < R_{200}$ from the cluster.

This measurement was done for passive and star-forming galaxies separately. The segregation between the two populations is done using the SED classification given by {\it LePhare} at the galaxy best photo-$z$. We also split our cluster sample in two redshift bins $0.15 < z \le 0.4$ and $0.4 < z < 0.7$ respectively. We note that the classification of galaxies as passive or star-forming using SED fitting with five optical bands is not extremely robust for individual galaxies, as the star-formation rate of galaxies cannot be computed precisely. This may introduce some noise in our measurements.

In the following, we compare the distribution of $d_{\rm
  clus,fil}$ for the two galaxy populations (passive and star-forming) and redshift bins (low redshift and high redshift). To this aim we give the estimated medians of each distribution and compare
them.
The formal comparison of the distributions was done by performing a
Kolmogorov-Smirnov (KS) test on the CDFs with 10000 bootstraps. For
each bootstrap realisation we test the null
hypothesis that the samples are drawn from the same parent
distribution. The significance of the difference between the samples
is then quantified through the fraction of bootstrap realisations for which
the null hypothesis is rejected (KS $p-{\rm value} < 0.01$).

\indent Results are presented in Fig.~\ref{fig:dclus_z_type}, where we show the
distributions of $d_{\rm clus,fil}$ for passive galaxies and star-forming galaxies in the low
(left) and high (right) redshift bins and in Fig.~\ref{fig:dclus_type_z}, where we show the
distributions of $d_{\rm clus,fil}$ for passive galaxies (left) and star-forming galaxies (right)
in the two redshift bins. Note that in these two figures, we are
displaying the same four distributions but Fig.~\ref{fig:dclus_z_type}
focuses on the galaxy-type difference at a given redshift, while
Fig.~\ref{fig:dclus_type_z} focuses on the redshift evolution for a
given galaxy type.

\indent We see in Fig.~\ref{fig:dclus_z_type} that there is no difference in the distance distribution
between passive galaxies and star-forming galaxies at high
redshift (the KS null hypothesis {\it cannot} be rejected in $\sim 95\%$ of
bootstrap resamplings, see Fig.~\ref{fig:KS-zh}).
On the other hand, passive galaxies are slightly closer
to clusters at low redshift (the KS null hypothesis {\it can} be rejected in $\sim 90\%$ of
bootstrap resamplings, see Fig.~\ref{fig:KS-zl}). If interpreted in the context of the colour-density relation
\citep[e.g.][]{Cooper2007}, the difference at low redshift actually confirms that by applying
\disperse~with photo-$z$s we are able to recover a density increase
along filaments up to {\small AMASCFI} outside $R_{200}$.

When looking at Fig.~\ref{fig:dclus_type_z}, we see that both galaxy
populations see their distances to clusters along filaments
decrease with decreasing redshift, but the trend is stronger
for passive galaxies. The difference between the distributions is
significant in both cases (the KS null hypothesis {\it can} be rejected in $\sim 98\%$ of
bootstrap resamplings, see Fig.~\ref{fig:KS-type}).

Values and differences of the distribution medians are reported in Table~\ref{tab:medians_type_z} where:
\begin{equation}
    \Delta_{X-Y} = \frac{\left \vert d_X - d_Y\right \vert}{\sqrt{\sigma_{d_X}^2 + \sigma_{d_Y}^2}},
    \label{eq:Delta}
\end{equation}
whith $d_X$ and $d_Y$ the distances (or their medians) and $\sigma_{d_X}$ and $\sigma_{d_Y}$ the associated uncertainties. These results are discussed in Sect.~\ref{sec:discussion}.

\begin{figure}
  \centering 
  \includegraphics[width=0.45\textwidth]{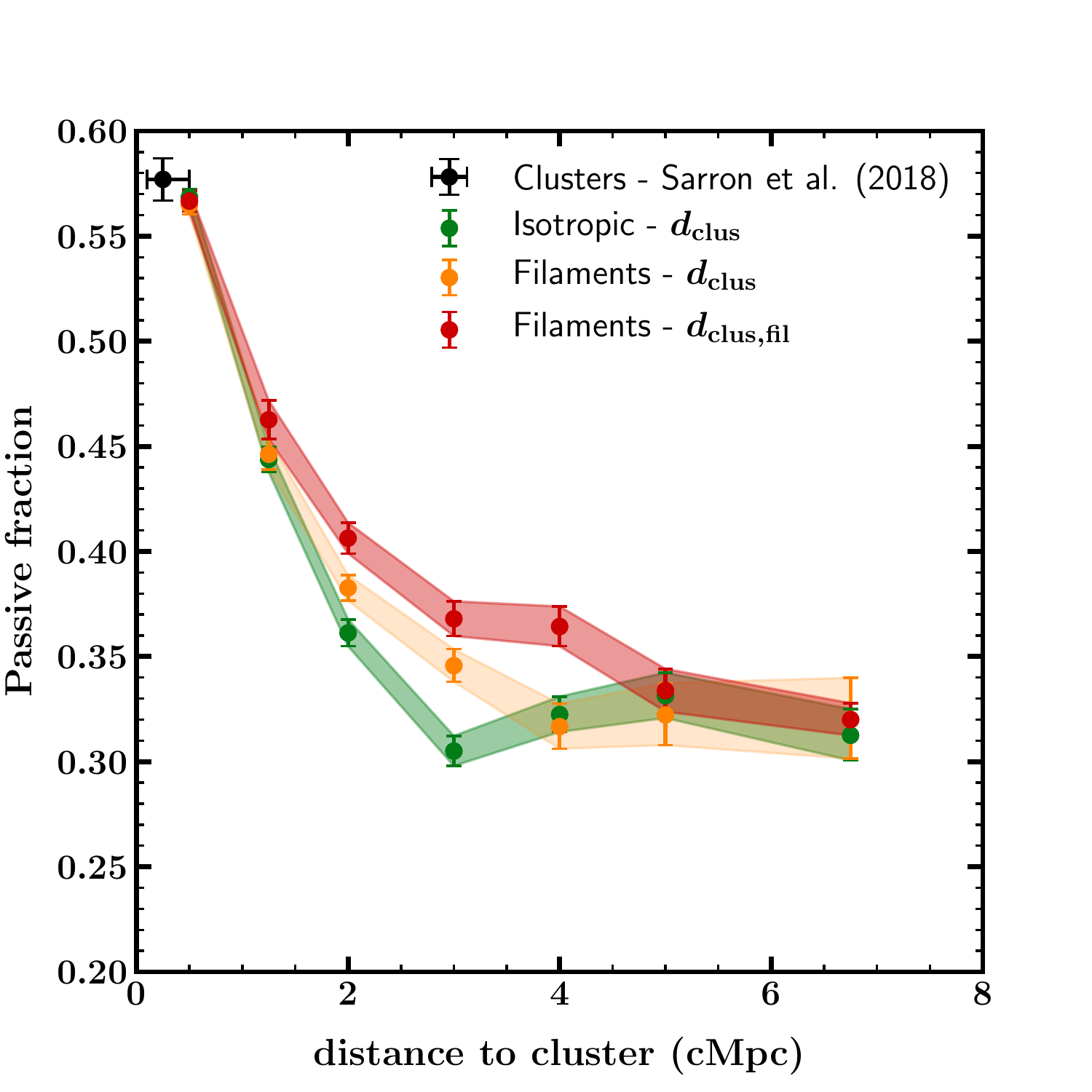}
  \caption{Galaxy passive fraction as a function of the
    distance to the cluster in comoving Mpc. Red points are
    galaxies in filaments connected to the clusters, whose
    clustercentric distance is computed along the filaments. The
    results for the same galaxy sample but with clustercentric
    cartesian distance is plotted in yellow. In green we plot
    galaxies in annuli around the clusters, considered to fall on the cluster isotropically.}
  \label{fig:pfrac}
\end{figure}

\subsection{Passive fraction}\label{sec:pfrac}
If quenching is efficient in cosmic filaments, we might expect to see an increased fraction of passive galaxies in their galaxy population compared to the field. To check for this, we computed the passive fraction in filament galaxies. Filament galaxies were selected and split between passive and star-forming galaxies as in Sect.~\ref{sec:dclus_CFHTLS}.

To ensure that we probe the specific effect of cosmic filaments on the
passive fraction, we proceeded in a similar way to
\citet{Martinez2016}. We chose as a reference sample galaxies in the
infall region of clusters ($d_{clus} > R_{200}$) but outside filaments.
For each cluster, we selected galaxies whose closest projected node in
the slice is the cluster and that are not located in filaments
($d_{\rm skel} > 1$cMpc). As in \citet{Martinez2016}, we refer to these as isotropically falling galaxies or isotropic galaxies. The passive fraction in filaments is found to be $f_{\rm passive,fil} = 0.370 \pm 0.006$ while the passive fraction of isotropic galaxies is $f_{\rm passive,iso} = 0.301 \pm 0.007$.

Then, we computed the passive fraction of filament galaxies as a function of their distance to the cluster over the full redshift range $0.15 < z < 0.7$. Results are shown in red in Fig.~\ref{fig:pfrac}. Error
bars and shaded areas are the 68\% confidence intervals for binomial population proportions computed following \citet{Cameron2011}. The passive fraction of isotropic galaxies as a function of distance to the cluster is plotted in green.

When we compute the distance to the cluster along the filaments
$d_{\rm clus,fil}$, we may introduce a bias compared to the
isotropic sample, as filaments wind around on their way to the cluster. Thus, to
cancel out this effect, we compute the cartesian distance to the cluster
for galaxies in filaments (the yellow points in Fig.~\ref{fig:pfrac}).

Both for filament galaxies and isotropic galaxies, there is a smooth
decrease of the passive fraction as a function of increasing clustercentric
distance. The value at $d_{clus} < 1$ cMpc is compatible with the
passive fraction observed in clusters by \citet{Sarron2018}.

The passive fraction remains higher in filaments as the distance from
the cluster increases up to $d_{clus} \sim 4-5$ cMpc, which roughly corresponds to $(2.5-3) R_{200}$.  The excess
of passive galaxies remains between $2$ and $4$ cMpc when considering the
cartesian distance to the cluster. These results are discussed in Sect.~\ref{sec:discussion}.

\begin{figure}[h!]
  \begin{subfigure}[t]{0.5\textwidth}     \includegraphics[width=1.\textwidth]{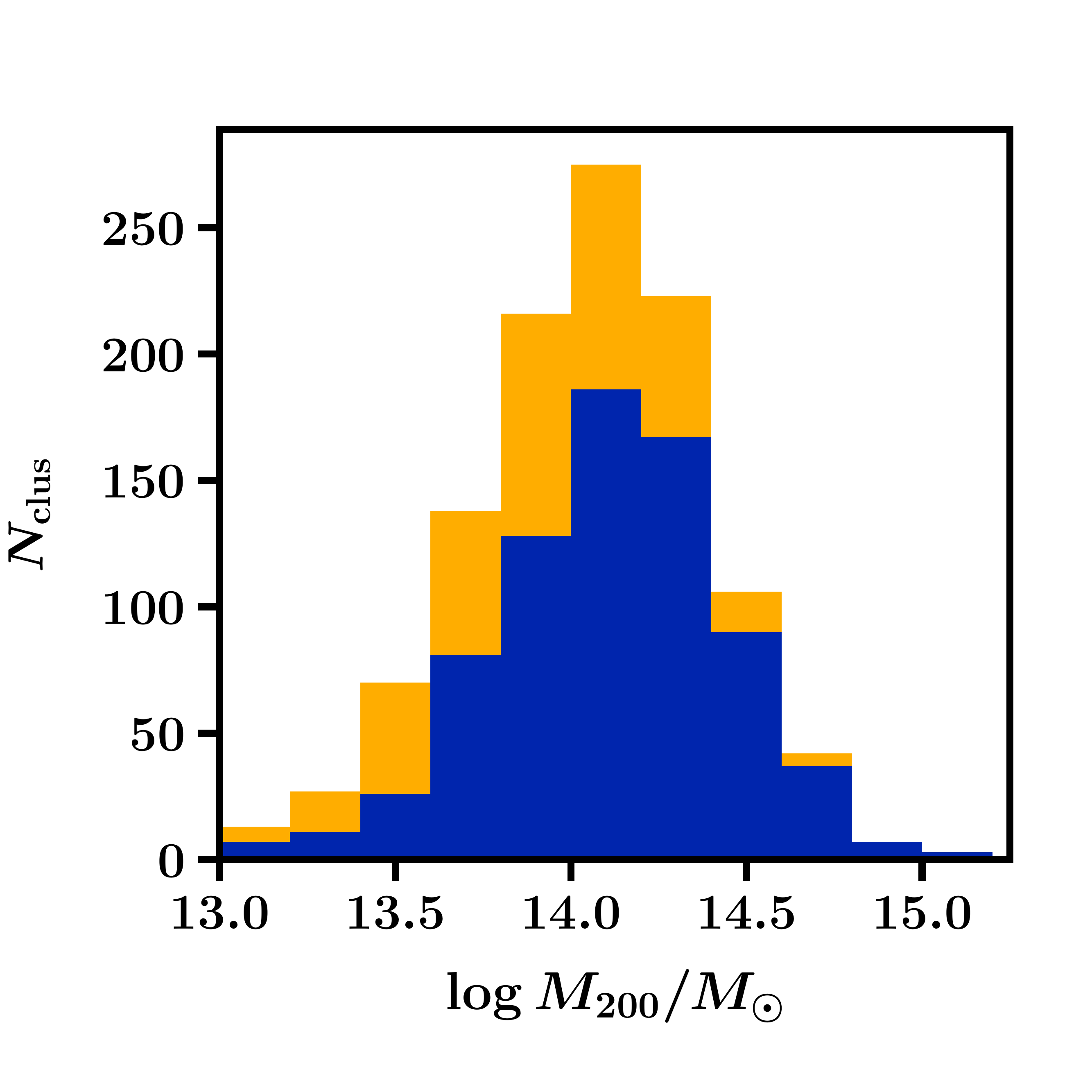}
    \label{fig:unmatched_halos}
  \end{subfigure}
  \begin{subfigure}[t]{0.5\textwidth}
    \hspace*{0.25cm}
\includegraphics[width=1.\textwidth]{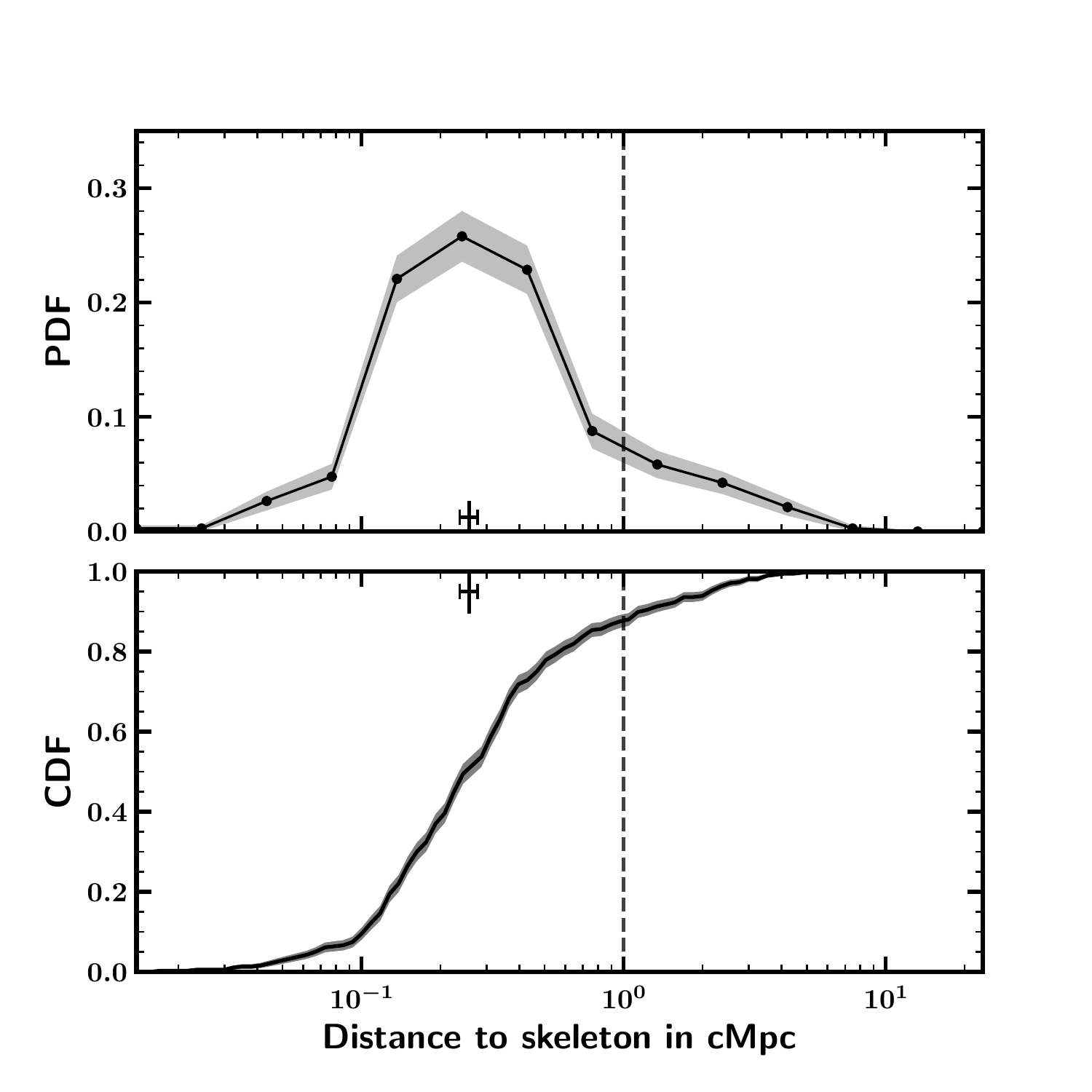}
    \label{fig:dgrp_to_clus_W1}
  \end{subfigure}
  \caption{Top: histogram of 
      groups/clusters respectively matched (blue) and unmatched (gold) with \disperse~nodes, as a function of their $M_{200}$ in the CFHTLS-W1 field. The two
      histograms are stacked on top of each other so that their sum is
      the total number of groups/clusters in the mass bin. Bottom: distribution of the distances to the skeleton of groups not
    located at a node in CFHTLS-W1. The dashed line is the chosen radius for filaments ($r = 1$ cMpc). See Fig.~\ref{fig:dclus_z_type} for details on symbols.}
  \label{fig:dgrp_clus}
\end{figure}

\subsection{Galaxy groups inside filaments}\label{sec:dgroup_fil}
As already mentioned, according to
\citet{Libeskind2018}, on large scales most filament finders tend to locate groups of
$M_{200} \sim 10^{13.5}$~M$_\odot$ inside cosmic filaments rather than at the nodes of the cosmic web. If groups are indeed located in filaments, then they may play a role in the observed gradient towards clusters observed in  Sect.~\ref{sec:dclus_CFHTLS} and in the passive fraction observed in Sect.~\ref{sec:pfrac}.

As mentioned in Sect~\ref{sec:connect}, when computing the
connectivity, some halos are not matched to \disperse~nodes. The
proportion of unmatched halos is actually a function of halo
mass. This is shown in Fig.~\ref{fig:dgrp_clus} where we plot the
histogram of matched and unmatched halos as a function of $\log
M_{200} / M_\odot$ in CFHTLS-W1. We see that most massive clusters are
all matched to a node from the skeleton reconstruction. Yet many low
mass {\small AMASCFI} candidate clusters and groups are not matched to
a node.

We thus computed the distances of these unmatched groups to
the filaments traced by \disperse~in order to see where they are located
with respect to the filaments of the cosmic web. The distances are computed in two
dimensions, in the slice centred at the cluster redshift.

Results are displayed in Fig.~\ref{fig:dgrp_clus}. Looking at the
CDF, we see that more than $75\%$ of groups are located in filaments ($d_{\rm skel} < 1 \
{\rm cMpc}$). These results have implications on the interpretation we can give to
the galaxy-type gradient towards clusters along filaments that we observed
in Sect.~\ref{sec:dclus_CFHTLS} and to the observed passive fraction in
filaments observed in  Sect.~\ref{sec:pfrac}, as discussed in Sect.~\ref{sec:discussion}.
 
\section{Discussion}\label{sec:discussion}

In Sect.~\ref{sec:dclus_CFHTLS}, we
showed that the median distance to clusters along filaments of both
passive and star-forming galaxies is slightly higher at high
redshift than at low redshift. While the trend is faint, a KS test confirmed that it is significant (see Figs.~\ref{fig:KS-z} and \ref{fig:KS-type}).
This would agree with a picture where galaxies
follow filaments towards the cluster potential well in the redshift range $0.15 < z < 0.7$.

When comparing the distributions of passive and star-forming galaxies in the two redshift bins separately (see Fig.~\ref{fig:dclus_z_type}), we observed a galaxy type gradient towards clusters inside filaments in the low redshift bin only ($0.15 < z_{clus} \le 0.4$), passive galaxies being located in the regions closer to clusters along filaments than star-forming galaxies. At high redshift ($0.4 < z_{clus} < 0.7$), the distributions are
the same for passive galaxies and star-forming galaxies.
So passive galaxies in filaments are located closer to clusters in the low redshift bin but not in the high redshift bin. Again, the trend, while faint, was confirmed to be significant by a KS test.

Moreover, we showed in Sect.~\ref{sec:pfrac} that over the full
redshift range ($0.15 < z < 0.7$), the fraction of passive galaxies in
filaments is higher than in regions around
clusters outside of filaments (that we referred to as isotropic regions), and the fraction of passive galaxies in filaments decreases with increasing clustercentric distance. This agrees with the findings of
\citet{Martinez2016} at $z < 0.15$ and \citet{Salerno2019} at $0.43 \le z \le 0.89$, while exploring redshifts intermediate between these two studies. This is also in agreement with the findings of \citet{Kraljic2018} in the GAMA survey in the range $0.03 \le z \le 0.25$ that found that the red fraction depends simultaneously on the distance to the filament and the distance to nodes. \\
\indent We remind the reader that our results are based on classification of passive and star-forming galaxies using SED fitting. This classification is statistically correct but is not extremely robust for individual galaxies when using five optical bands as it is the case in the CFHTLS. Some galaxies may thus be wrongly classified as passive or star-forming, introducing noise in our galaxy-type gradient and passive fraction measurements.

Finally, we showed in Sect.~\ref{sec:dgroup_fil} that some low mass clusters and groups
in the AMASCFI catalogue are not located at nodes as detected by our
skeleton reconstruction. Looking at the distances of these unmatched groups to the skeleton, we
found that most of them ($\sim 80\%$) are located in filaments.  This
is in agreement with the 
scenario of hierarchical structure formation, in which clusters keep 
accreting smaller groups in the redshift range $0.15 < z < 0.7$
\citep[e.g.][]{Contini2016}.

These results can be interpreted in two ways. First, one can assume that
passive galaxies fall faster in the potential well that star-forming
galaxies. This would explain why passive galaxies are closer to
clusters than star-forming ones in the low redshift bin. The increased falling speed could come from a higher radial speed along the filament due to the fact that passive galaxies are located closer to the filament spine \citep{Laigle2018,Kraljic2018}. One way to
understand this is also by thinking of passive galaxies being located preferentially in
galaxy groups, that we showed to be mainly located closer to the filament spine and may thus fall faster onto clusters than isolated
galaxies. In this case, the fact that we do not observe a significant
difference in the distance distributions of passive and star-forming
galaxies in our high redshift bin would be coherent with a picture
where most of the collapse of groups into clusters occured at $z < 0.7$ \citep{Contini2016}.

On the other hand, the results can also be interpreted by quenching
occurring in the filaments. This would require that the quenching
mechanism is more efficient when galaxies in the filament are located
closer to the cluster. This could fit with a scenario where during their
journey along the filaments, galaxies have a higher probability to collide or
enter a group as satellites, and even more so when they get closer to the
cluster. Both phenomena (merger and group accretion) are known to be
responsible for quenching. Moreover, in our redshift range, filaments
keep accreting field galaxies from the walls/voids in the saddle point
regions. As most of these galaxies should be star-forming \citep[e.g.][]{Kraljic2019},
this would tend to increase the median distance of
star-forming galaxies to clusters.

In this second scenario, even though our results cannot assess which physical processes might be
at play in the filament quenching, it is interesting
to note that we also showed that we found groups located close to the filament spines. This could point towards filament quenching being due to
pre-processing by galaxy groups through strangulation as argued by
\citet{DeLucia2012} or \citet{Peng2015}. Yet we lack conclusive evidence to draw firm
conclusions and this calls for further investigation.


\section{Conclusions}

We presented a method to detect the cosmic web filaments based on
photometric redshifts. We showed the ability of the method, already
applied to the COSMOS field by \citet{Laigle2018}, to statistically
reconstruct the filament distribution. In particular, we focused here on the infall regions around clusters, where we showed that the scaling of connectivity with cluster mass is recovered, and found a completeness of $\sim 70\%$ and a purity of $\sim 66\%$ for the connected filament reconstruction.

We then applied the method to the CFHTLS T0007 data to study filaments of galaxies around {\small AMASCFI} clusters \citep{Sarron2018}. For each cluster, we analysed the cosmic filaments connected to the clusters.

Studying galaxy properties in these filaments connected to clusters,
we find that galaxies are located closer to clusters in the redshift range $0.15 < z \le 0.4$ compared to $0.4 < z < 0.7$. In the low redshift bin, we observed a galaxy-type gradient towards clusters, i.e. passive galaxies are located closer to clusters than
star-forming galaxies. Such a gradient does not exist in our high
redshift bin. In the full redshift range, we showed that the
fraction of passive galaxies is higher in our filaments than in
istropically selected regions around clusters and that the passive
fraction in filaments decreases with increasing distance to the cluster up to $d_{\rm clus} \sim 5$ cMpc.

We proposed that this could be interpreted as quenching occuring in the
filaments before galaxies reach the cluster virial radius - so-called
pre-processing.
As we found that a large fraction of groups not located at the nodes of the reconstructed cosmic web are in fact inside filaments ($80 \%$ of groups at $d_{{\rm skel}} < 1$ cMpc), we postulated that this
pre-processing could occur in galaxy groups
during the hierarchical growth of structures in agreement with
previously proposed quenching models \citep[e.g.][]{Wetzel2013,Moutard2018}.

We plan on building upon this proof-of-concept study by stuying the
passive and star-forming galaxy luminosity functions (GLF) of our
filaments. This will enable us to better interpret the trends observed in
the cluster GLFs and to pinpoint the physical processes at play in
quenching star-formation in dense environments.

This method is promising as it uses photometric redshifts of accuracy
typical of what is expected for future wide surveys such as {\it
  Euclid} and {\small LSST}, that will allow to explore cosmic
filaments at even higher redshifts with more statistics.

\bibliographystyle{aa}
\bibliography{sarron_biblio}

\begin{appendix}
  \section{KS tests on distance distributions}
  As mentioned in Sect.~\ref{sec:dclus_CFHTLS}, to check if the
  distributions of distances to clusters of galaxies in filaments
  are different for passive and star-forming galaxies in the low ($0.15 < z \le 
  0.4$) and high ($0.4 < z < 0.7$) redshift bins, we performed
  Kolmogorov-Smirnoff (KS) tests on 10000 bootstrap realisations of the
  samples. The same was done for passive and star forming galaxies.

\begin{figure}[h]
  \begin{subfigure}[t]{0.5\textwidth} 
   \hspace*{-0.25cm} \includegraphics[width=1.\textwidth]{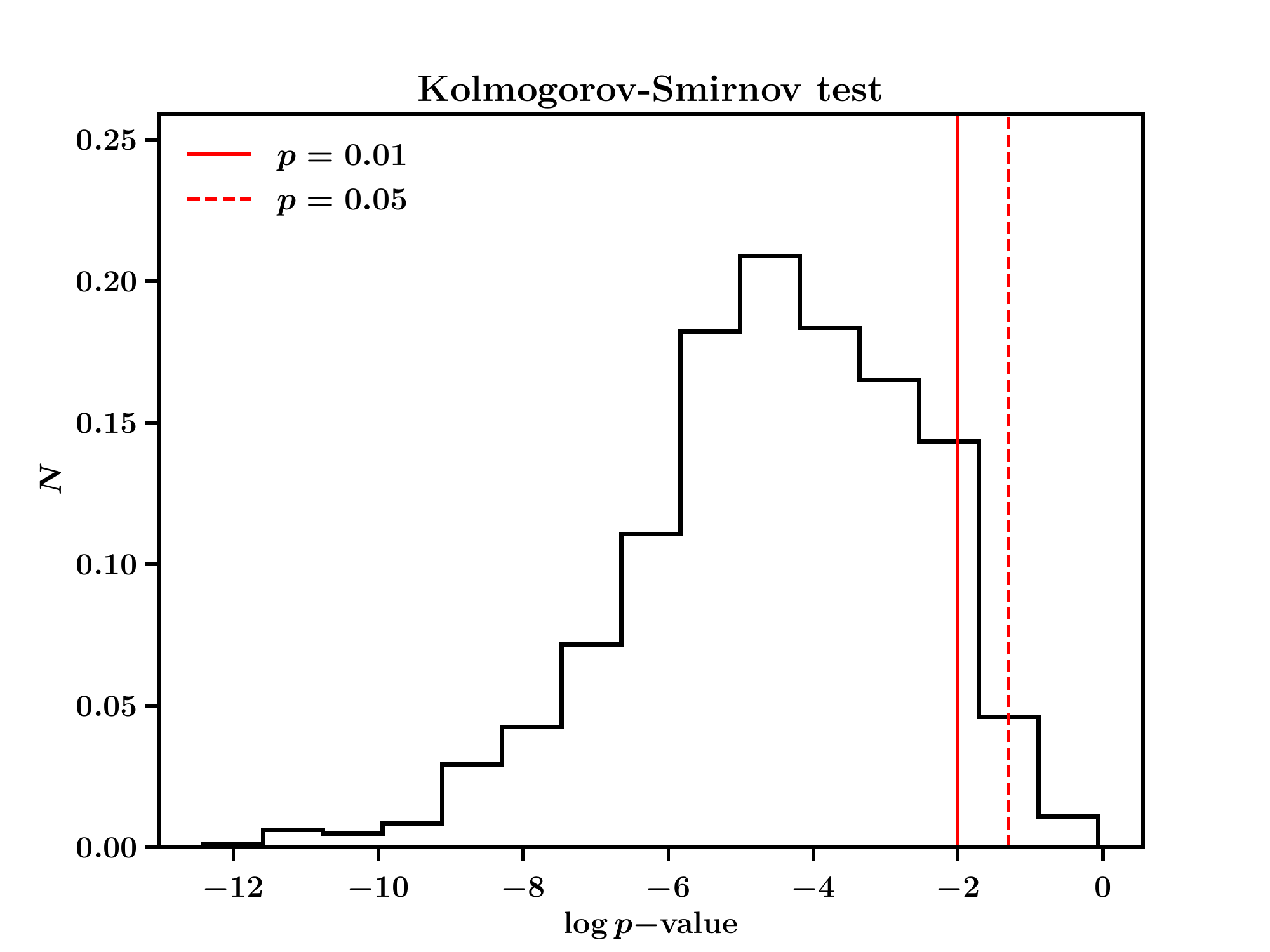}
    \caption{$0.15 < z < 0.4$}
    \label{fig:KS-zl}
  \end{subfigure}
  \begin{subfigure}[t]{0.5\textwidth}  
  \hspace*{-0.25cm} \includegraphics[width=1.\textwidth]{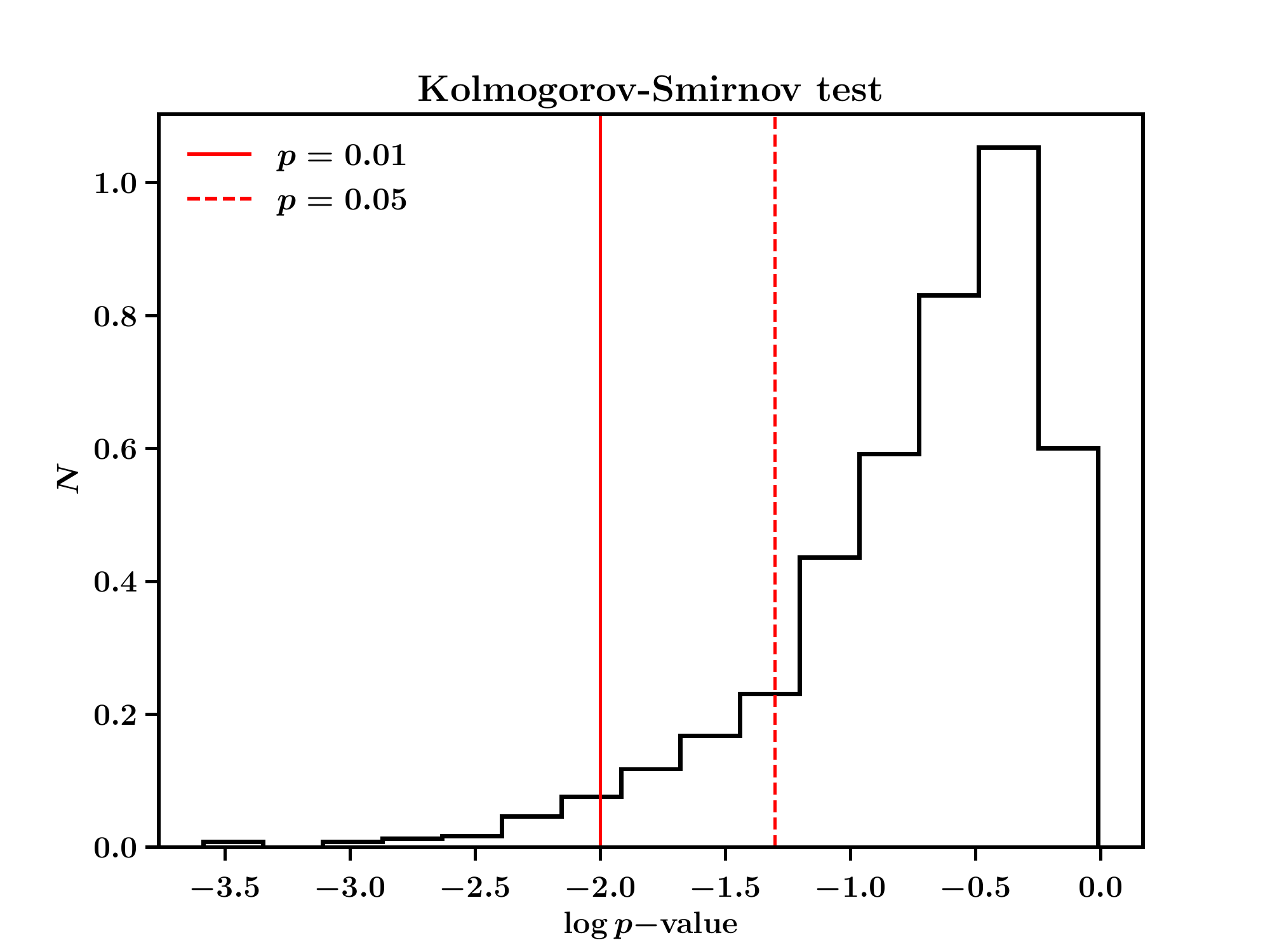}
    \caption{$0.4 < z < 0.7$}
    \label{fig:KS-zh}
  \end{subfigure}
  \caption{$p-$value histograms of 10000 bootstrap realisations of KS
    tests on the distributions of Fig.~\ref{fig:dclus_z_type} at low (a)
    and high (b) redshifts respectively.}
  \label{fig:KS-z}
\end{figure}

\begin{figure}[h]
  \begin{subfigure}[t]{0.52\textwidth} 
   \hspace*{-0.25cm} \includegraphics[width=1.\textwidth]{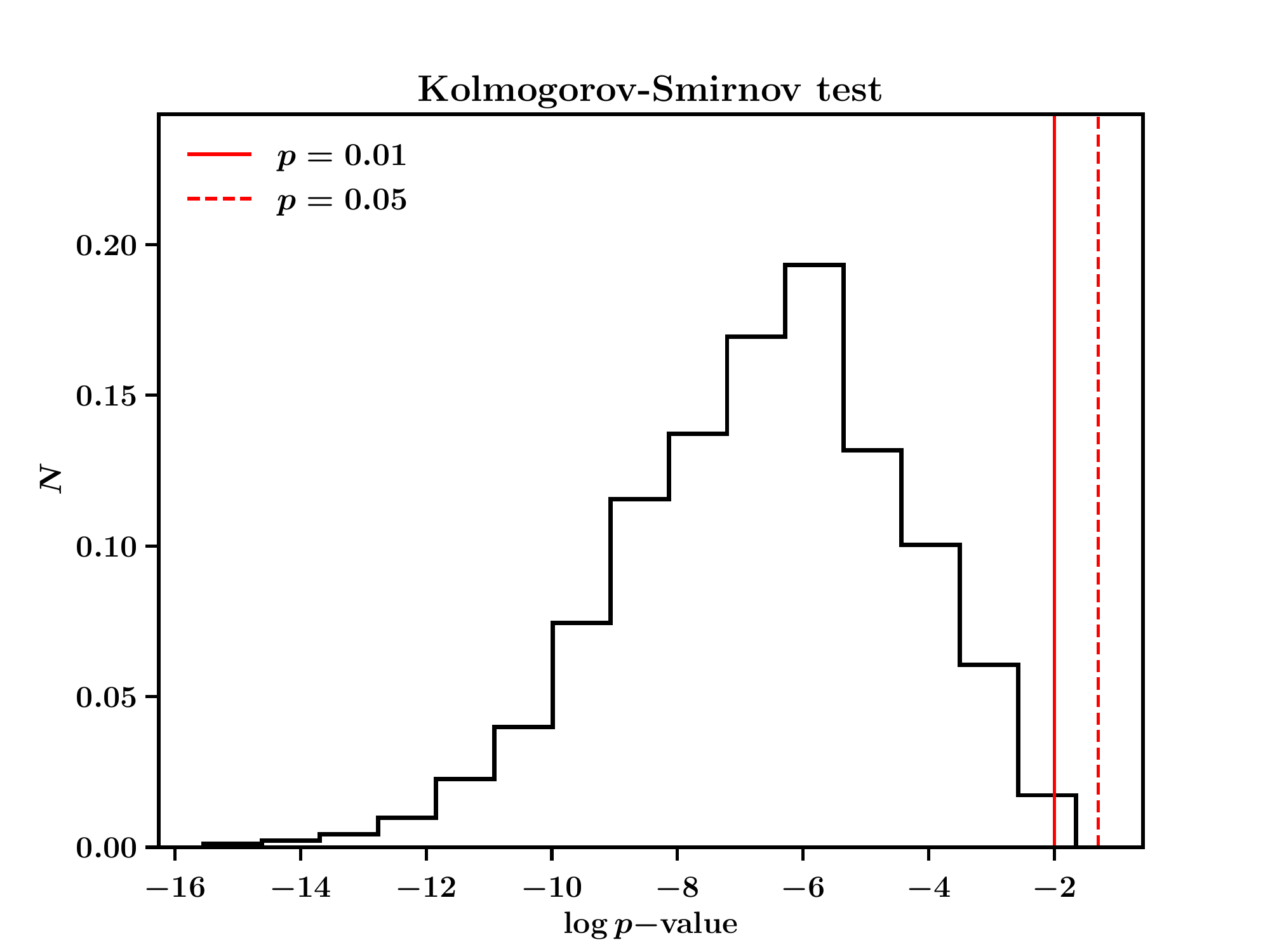}
    \caption{passive galaxies}
    \label{fig:KS-early}
  \end{subfigure}
  \begin{subfigure}[t]{0.52\textwidth}  
  \hspace*{-0.25cm} \includegraphics[width=1.\textwidth]{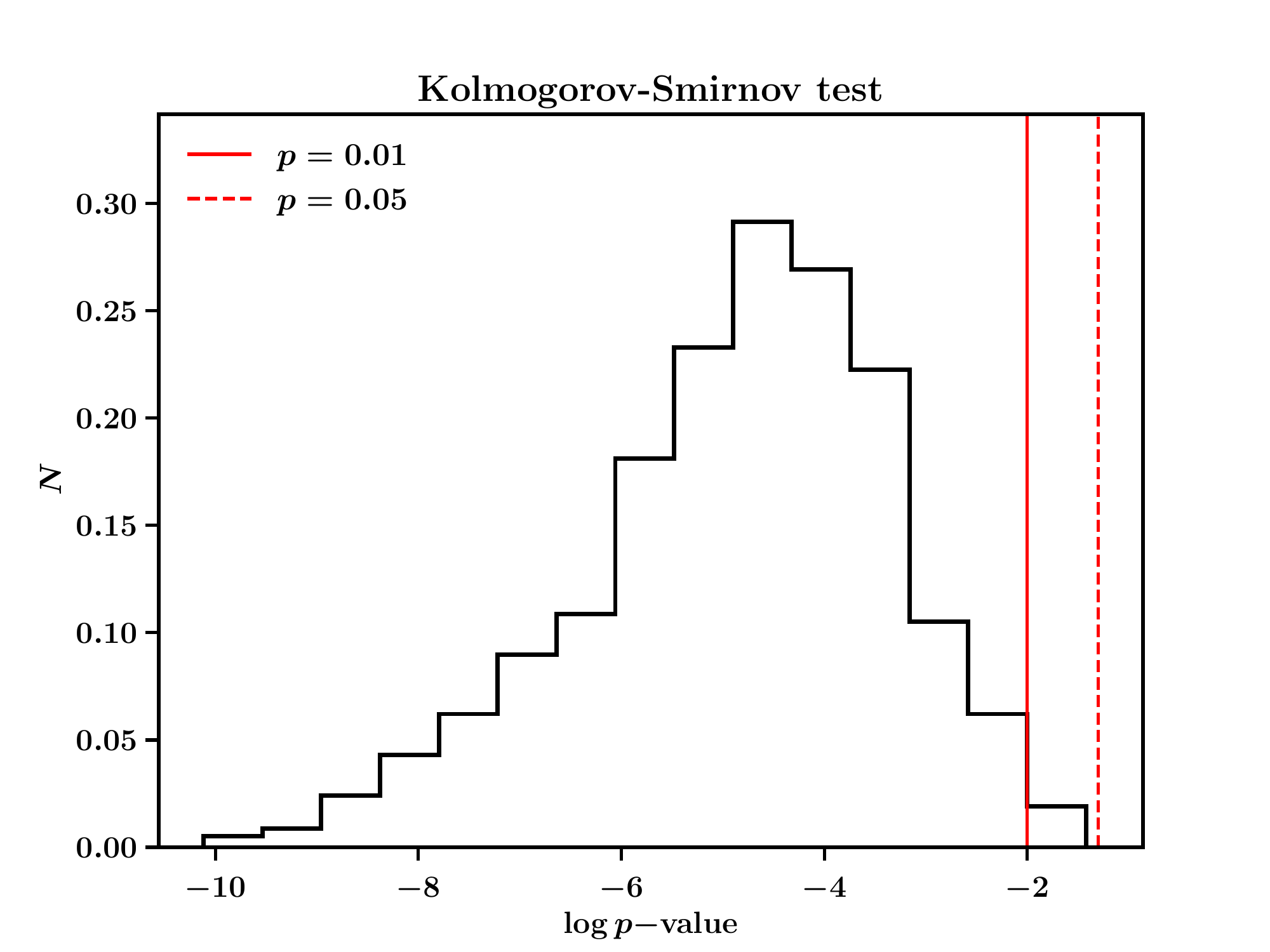}
    \caption{star-forming galaxies}
    \label{fig:KS-late}
  \end{subfigure}
  \caption{$p-$value histograms of 10000 bootstrap realisations of KS
    tests on the distributions of Fig.~\ref{fig:dclus_type_z} for passive (a)
    and star-forming (b) galaxies respectively.}
  \label{fig:KS-type}
\end{figure}

  The distributions of the $p-$values of these KS tests are presented
   in Figs.~\ref{fig:KS-z} and Fig.~\ref{fig:KS-type}.

\end{appendix}

\end{document}